# A Needle in a Haystack – How to Derive Relevant Scenarios for Testing Automated Driving Systems in Urban Areas

Nico **Weber**, M. Sc.; Dr.-Ing. Christoph **Thiem**
Opel Automobile GmbH, Stellantis, 65423 Rüsselsheim am Main, Germany

Prof. Dr.-Ing. Ulrich **Konigorski**
Control Systems and Mechatronics Laboratory (rtm), TU Darmstadt, 64283 Darmstadt, Germany

Contact: nico.weber@external.stellantis.com

**Summary**

While there was great progress regarding the technology and its implementation for vehicles equipped with automated driving systems (ADS), the problem of how to proof their safety as a necessary precondition prior to market launch remains unsolved. One promising solution are scenario-based test approaches; however, there is no commonly accepted way of how to systematically generate and extract the set of relevant scenarios to be tested to sufficiently capture the real-world traffic dynamics, especially for urban operational design domains. Within the scope of this paper, the overall concept of a novel simulation-based toolchain for the development and testing of ADS-equipped vehicles in urban environments is presented. Based on previous work regarding highway environments, the developed novel enhancements aim at empowering the toolchain to be able to deal with the increased complexity due to the more complex road networks with multi-modal interactions of various traffic participants. Based on derived requirements, a thorough explanation of different modules constituting the toolchain is given, showing first results and identified research gaps, respectively. A closer look is taken on two use cases made possible by the toolchain. First, it is investigated whether the toolchain is capable to serve as synthetic data source within the development phase of ADS-equipped vehicles to enrich a scenario database in terms of extent, complexity and impacts of different what-if-scenarios for future mixed traffic. Second – inspired by concepts of present works – it is analyzed how to combine the individual advantages of real recorded data and an agent-based simulation within a so-called adaptive replay-to-sim approach to support the testing phase of an ADS-equipped vehicle. The developed toolchain contributes to the overarching goal of a commonly accepted methodology for the validation and safety proof of ADS-equipped vehicles, especially in urban environments. In particular, it supports the achievement of the answer concerning a holistic methodology for the generation and extraction of relevant scenarios to be tested in urban environments tackling the open-context problem. Besides, the combination of a real-data based scenario extraction with agent-based simulation approaches can contribute to find previously unknown unsafe scenarios for an ADS-equipped vehicle.

**Content**

## 1   Introduction

Proving the safe operation of vehicles equipped with automated driving systems (ADS) [1] prior to their market launch stays a yet unsolved problem. While there was great progress regarding perception technology and computational power in the last decades, questions concerning another necessary condition for the mass deployment of ADS-equipped vehicles, namely a commonly accepted methodology for the validation and safety proof, still remain unanswered [2], [3], [4].

### 1.1   Urban Environments in Focus

Research and development into these systems is paid particular attention by various stakeholders ranging from industry and academic institutions over regulatory authorities and insurance companies to different parts of society [5]. The main reasons for this vast interest are the potentials of ADS-equipped vehicles to contribute significantly to a safer, more efficient and more comfortable future mobility [6], [7]. Within many research projects and publications in the last years, investigations focused on a highway environment as operational design domain (ODD) [1], whilst less research was conducted regarding urban environments [8], [9]. However, on average, in the last five years (2016-2020), nearly 70 % of all reported accidents involving personal injuries in Germany occurred in urban areas according to [10]. It is of note that in about 2/3 of all accidents in this period misbehavior by the driver was observed [10]. Moreover, business models like mobility as a service or autonomous delivery systems seem to be particularly interesting within urban areas [11].

## 1.2 New Challenges through ADS-equipped Vehicles

The fundamental challenge of proving safety for ADS-equipped vehicles can be traced back to the change in responsibility distribution between driver/user and vehicle in the case of automated driving (SAE-Level ≥ 3 according to [1]). For an execution of the dynamic driving task by the ADS in a predefined ODD, a validation and safety proof of the ADS-equipped vehicle and the intended functionality of the ADS has to be achieved [12]. This means that on the one hand it has to be proven that there occurs no unreasonable residual risk due to systematic failures or random hardware faults during the operation of the ADS-equipped vehicle, e.g., malfunctions of the E/E system [13]. On the other hand, according to ISO/PAS 21448, the absence of unreasonable risk caused by the intended functionality or performance limitations of the ADS has to be ensured, what is referred to as safety of the intended functionality (SOTIF) [14] and in the main focus of this paper.

This is challenging since existing approaches such as the statistical, distance-based proof of safety would require billions of test kilometers under representative conditions before market launch [15]. Therefore, new methods for the release of ADS-equipped vehicles are currently under development. One of these methods is the so-called scenario-based approach, as proposed by project PEGASUS [16]. It is based on the assumption that a reduction of the test effort is to be expected when testing exclusively relevant scenarios.

As scenarios on different abstraction levels [17] are a key factor within the scenario-based test approach, a lot of research has already been conducted regarding their feasible generation and extraction within a highway environment [18], [19], [20]. However, even when applying the scenario-based approach it comes with the inevitable problem of the open context (also known as open-world challenge) within which the ADS-equipped vehicle has to operate safely. Since the proof of safety cannot be performed by exclusively real-world testing in this open context, virtual testing approaches must be considered as crucial pillar within scenario-based testing approaches. Thereby, simulation-based methods are able to meet vital requirements regarding a test case execution in a safe and efficient manner, but come inevitably with the intrinsic challenge of verification and validation of the simulation tools and results themselves [12], [14], [21].

According to ISO/PAS 21448, scenarios occurring in this open context can intellectually be divided in four groups – known safe, known unsafe, unknown unsafe and unknown safe scenarios, as shown in Fig. 1 (left). For a toolchain, which covers the generation and extraction of relevant scenarios there are two primary goals. First, the known unsafe scenarios, which can be explicitly evaluated, have to be provided for subsequent testing of the ADS-equipped vehicle. Second, the toolchain has to offer the ability to reveal the before unknown unsafe scenarios for the ADS-under-test. Through this, an argument can be provided that these two scenario groups are sufficiently small for the ADS-under-test and therefore that the resulting residual risk is acceptable [14], as shown in Fig. 1 (right). In other words, the infinite-dimensional open context has to be mapped to a finite and manageable set of scenarios [22] that reflects the nature of the traffic dynamics of interest in a sufficiently valid manner for subsequent testing.

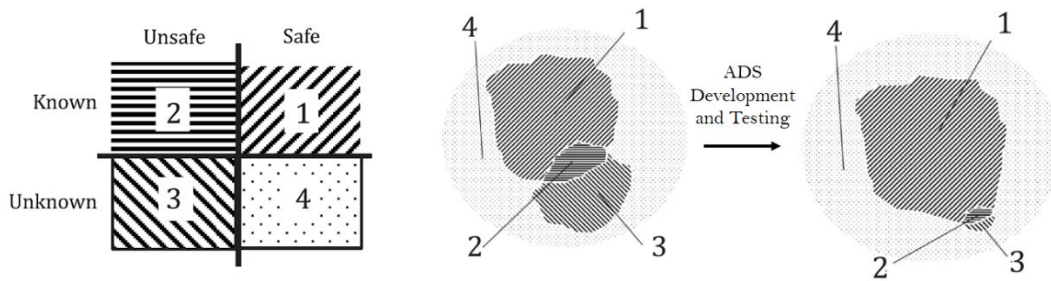

Fig. 1    Different scenario categories (left) and scenario category evolution (right) based on [14]

## 1.3    Tackling the Open-context Problem

To the best knowledge of the authors, there is no commonly accepted methodology of how to reveal the aforementioned unknown, unsafe scenarios to be tested systematically, especially for urban ODDs. The majority of present work regarding data-driven scenario extraction is conducted within a highway environment [2], [18], [19], [20], [23]. Since the multi-modal interaction of various traffic participants is characterizing the traffic dynamics within urban areas, the toolchain presented in this work shall be particularly able to extract scenarios based on trajectory data, e.g., recorded by unmanned aerial vehicles (drones). This kind of data source provides the opportunity of an objective view of a scene [24] and, hence, to capture the relevant traffic dynamics surrounding the ADS-under-test [25], [26]. Publications dealing with an urban ODD mostly focus on scenario extraction based on other data sources, e.g., accident data or CAN-based data of measurement vehicle, and particularly analyze vehicle-to-vehicle scenarios [27], [28], [29]. In terms of scenario extraction based on trajectory data, in the work of King et al. [30] an approach for deriving logical vehicle-to-vehicle scenarios for an unsignalized intersection in the U.S. is presented. In addition to real-data based scenario extraction, the toolchain presented in this paper shall be able to support the creation of a hybrid scenario database containing both real and synthetic data. Purely synthetic-data based scenario extraction approaches are presented in [31] for a highway environment and in [25] for urban environments. The mix of both real and synthetic data for scenario extraction is presented in [20] for a highway environment and in [28] for urban environments, whereby the focus of [28] is on maneuver classification for trajectory prediction. While Sippl [25] mentions the flexible and efficient data generation through simulation, Barbier et al. [28] state that a relatively small data set containing synthetic and real data is able to accurately classify real maneuvers. Additionally, they conclude that information from exteroceptive sensors could be helpful for further development. Although there is relatively little research done concerning the construction of a hybrid scenario database such an approach provides the potential to enrich a scenario database in terms of extent, complexity and impacts of different what-if-scenarios for future mixed traffic, what will be further elaborated in Section 4.

In literature, various simulation frameworks for the development and testing of ADS-equipped vehicles are presented [32], [33], [34]. The majority of the presented approaches deals with a highway environment and only a part of them is mainly focused on scenario generation and extraction [2], [35]. However, there is also work done regarding simulation frameworks for the development and testing of ADS-equipped vehicles in urban environments, e.g., in [9] or within the SET Level research project [36], a successor of the PEGASUS research project and part of the PEGASUS family [16].

## 1.4 Towards a Toolchain for the Development and Testing of ADS-equipped Vehicles in Urban Environments

Within the scope of this paper, the overall concept of a novel simulation-based toolchain for the development and testing of ADS-equipped vehicles in urban environments is presented. It applies the scenario-based test approach and is based on methodological concepts of previous works [2], [31]. The developed novel enhancements aim at empowering the toolchain to be able to deal with the increased complexity due to more complex road networks within urban environments, placing special emphasis on the multi-modal interaction of various traffic participants. Striving for a holistic approach in terms of scenario generation, extraction and execution, the paper investigates two different research questions. On the one hand, it is analyzed whether a toolchain can be built up, which is capable to serve as synthetic data source within the development phase to build up a hybrid scenario database containing both real and synthetic data. On the other hand, it is examined how to combine the individual advantages of real recorded data in form of trajectory data and an agent-based simulation within a so-called adaptive replay-to-sim approach within urban areas based on concepts described in [37], [38], [39]. This approach provides the possibility to support the testing phase of an ADS-equipped vehicle. It is characterized by replacing predefined trajectories of surrounding traffic participants by agent models while executing a concrete scenario in case of a significant deviation between the ADS-under-test and the human-driven vehicle in the real recorded scenario. This enables an interactive closed-loop simulation based on real recorded data, which can be enriched on different levels. Based on these two research questions, top-level requirements for the target toolchain are derived, which are subsequently transformed into a tool-agnostic functional system architecture. A thorough explanation of the different modules constituting the toolchain is given, whereby the core is represented by a coupled traffic and ego-vehicle simulation. Research gaps regarding different functional modules are revealed and emphasized based on first results of instantiated functional modules, respectively.

The main contributions of this paper can be summed up as follows:

1. Formulation of requirements for a simulation-based toolchain for the development and testing of ADS-equipped vehicles in urban environments (Sec. 3.1)

2. Derivation of a functional system architecture for the target toolchain (Sec. 3.2)
3. Investigation of research question one: Analysis regarding capability of the toolchain to serve as synthetic data source to enrich a scenario database (Sec. 4.3)
4. Investigation of research question two: Analysis regarding capability of the toolchain to apply an adaptive replay-to-sim approach (Sec. 4.1- 4.3)
5. Identification of research gaps within the field of the safety proof for ADS-equipped vehicles based on results of items 3 and 4 (Sec. 4 and Sec. 5)

## 2 Related Work

### 2.1 Simulation-based Frameworks / Toolchains for the Development and Testing of ADS-equipped Vehicles

Whilst there are various simulation-based frameworks for the development and testing of ADS-equipped vehicles presented in literature (cf. Sec. 1.3), the subset conducting in-depth investigations concerning the ability for deriving relevant scenarios or such focusing on urban environments are related to this work.

With respect to highway environments, Hallerbach et al. [2] present a generic simulation-based toolchain for the model-in-the-loop identification of critical scenarios for cooperative and automated vehicles. The core of the toolchain consists of a coupled vehicle dynamics, traffic and cooperation simulation. The methodology is illustrated by an instantiation of the toolchain in form of Eclipse SUMO for the traffic simulation representation and IPG CarMaker for the vehicle dynamics representation for different exemplary highway scenarios. The authors define safety and traffic quality metrics, which are merged for a binary classification whether a concrete scenario is critical or not. In [31] the toolchain is extended by a methodology for the automated derivation of concrete scenarios based on purely synthetic data, which serve as input for the aforementioned toolchain. Nalic et al. [35] introduce a virtual framework for testing and validation of ADS conceptually similar to the work of Hallerbach et al. [2]. The framework consists of a co-simulation between the microscopic traffic simulator PTV Vissim, a vehicle dynamics simulation in IPG CarMaker, a co-simulation controller and a post-processing module. Similar to the work of [31], they calibrate the microscopic traffic simulation based on real traffic measurements on a motorway section. This is accomplished by clustering the traffic data into seven clusters, which represent average traffic conditions in terms of typical traffic volume profiles. These seven clusters represent scenarios on an abstract level for the investigated motorway section through which the ego-vehicle with a Highway Chauffeur function is sent repeatedly to reveal critical scenarios. Nalic et al. [40] extend the methodology by a stress testing method with which it is possible to provoke defined maneuvers in the vicinity of an ego vehicle to force challenging situations for the ADS-under-test.

With respect to urban environments, Sippl et al. [9] present a distributed real-time simulation setup for automated driving function testing in urban areas. They state the necessity of a holistic simulation of urban traffic environments for the virtual testing of automated driving functions. The developed simulation framework consists of a coupled ego vehicle and microscopic traffic simulation as well as an additional pedestrian simulation. They extend the existing pedestrian behavior model on the tactical level and implement an extended behavior model on the operational level. A proof-of-concept for the simulation setup for an exemplary urban traffic scenario on a signalized intersection is performed. The authors state limitations due to the limited amount of data within the used dataset and want to increase the amount and quality of the implemented models in future work. A method for the identification of relevant traffic scenarios based on synthetic data of the presented framework is described in [25].

Moreover, there are ongoing research projects such as SET Level [36] and VVMethoden [41] as successors of the PEGASUS project [16], which deal with the question how to prove the safe operation of ADS-equipped vehicles in urban areas. While VVMethoden has its main focus on test strategies and methodologies, the SET Level project focusses mainly on the simulation-based development and testing of automated driving in urban areas. Concerning a holistic approach for deriving relevant scenarios in urban areas, the current status of the closed-loop traffic simulation for criticality analysis is of particular interest [42]. Thereby, an approach is presented how to explore the scenario space of a given logical scenario [17].

## 2.2 Scenarios in the Context of Development and Testing of ADS-equipped Vehicles

Within this paper, the term scenario is used in accordance to the definitions of [24] after which a scenario describes the temporal development between several scenes containing the scenery, dynamic elements and all actors' and observers' self-representations, extended by actions and events as well as goals and values to describe this temporal development. Scenarios can be divided into various abstraction levels. Whereas the level of abstraction decreases from the functional to the concrete scenario, the number of potentially relevant scenarios increases. Functional scenarios describe a scenario on a semantic level, whose entities and corresponding relations between those entities are described in a human-interpretable manner. Logical scenarios include the scenario description on a state space level, where the entities and relations among them are specified, e.g., by probability distributions of the corresponding parameters. A concrete scenario is derived by specifying concrete values for each parameter in the state space [17].

For structuring the entities constituting a scenario the PEGASUS project [16] proposes a 6-layer model, based on the works [43], [44], [45], [46]. In Fig. 2, the 6-layer model for structuring scenarios is shown with exemplary entities for each layer. According to this model, a scenario can be structured by descriptions of the road level, traffic infrastructure, temporary manipulations of the road level and traffic infrastructure, (movable) objects, environmental conditions and digital information. For further explanations regarding the 6-layer model, please refer to [47].

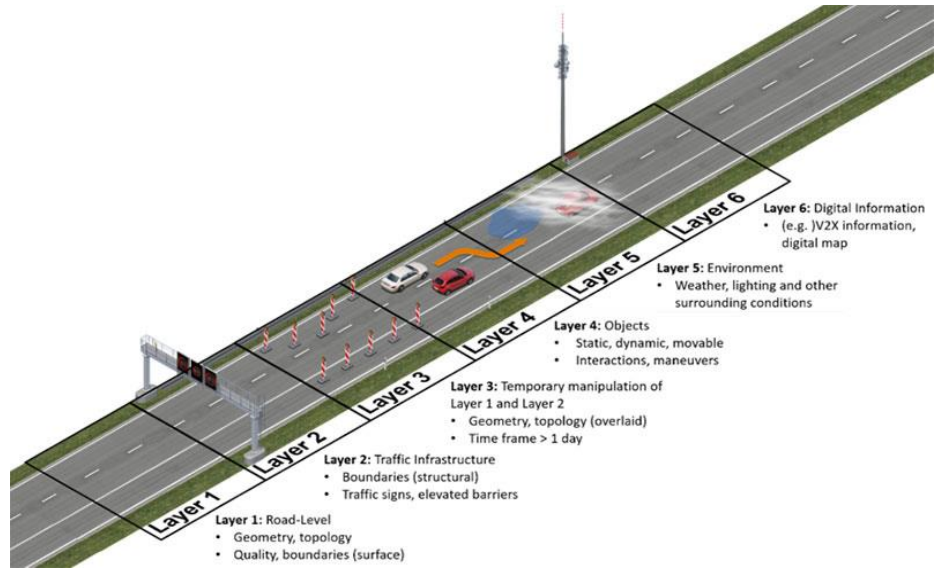

Fig. 2    6-layer model for structuring scenarios [16] based on [43], [44], [45], [46]

In general, scenarios for testing ADS-equipped vehicles can be extracted by applying a knowledge-based or a data-driven approach. The knowledge-based approach is characterized by the explicit definition of functional, logical or concrete scenarios by experts under use of abstract information like the functional specification of the test object, standards and guidelines, accident data etc. [43], [48]. For a structured representation of the resulting expert knowledge, ontologies can be used to finally derive scenarios [44]. Data-driven approaches are characterized by a data-mining process to reveal scenarios to be tested based on various data sources like field data, accident data or traffic simulation data [48], [49]. It has to be mentioned that these two approaches should not be interpreted as competitive but complementary [49]. While data-driven approaches might have advantages, e.g., regarding the scenario space coverage, knowledge-based approaches might be more suitable to identify edge cases, which are very rare to observe in real traffic but SOTIF-relevant for the ADS-under-test. Furthermore, it is noteworthy that, depending on the scenario types to be revealed, there are differences concerning the selection of a feasible data source when applying the data-driven approach. For a broad coverage of the scenario space field data might outperform accident data, while accident data might outperform field data when focusing on falsification experiments using corner-case scenarios. The problem of how to combine both the aforementioned approaches and various data sources to end up with a holistic, commonly accepted scenario database is yet unsolved. For a comprehensive overview regarding available scenario derivation methodologies, going beyond the explanations in the introduction, please refer to [48], [49].

## 2.3    Adaptive Replay-to-sim Approach

As mentioned in the introduction, the developed toolchain focusses on the answer of two different research questions at the current state of implementation. The capability of the toolchain to support activities on the left part of the V-model, including but not limited to scenario generation and extraction will be further elaborated in Section 4. A delimitation to existing publications is included in the introduction, while in Section 2.2 related work concerning scenarios in the context of ADS-equipped vehicle development and testing is described. In the following, emphasis is put on related work regarding the second research question, namely the implementation of a so-called adaptive replay-to-sim approach. This approach aims at combining the individual advantages of real recorded data and an agent-based simulation to support activities within the right part of the V-model (testing phase).

Based on [50] and [51] Wachenfeld and Winner [37] propose the concept of virtual assessment of automation in field operation (VAAFO), which is specified to some extent in [52]. Wang and Winner [38] introduce the concept in detail, which is generally based on the shadow mode (passive AD) validation method. Within this approach, an ADS-equipped vehicle is operated in a real traffic environment driven by a human. The automated driving function is provided with real inputs from the vehicle sensors and environmental sensors, but has no access to the real actuators of the vehicle. Through behavior comparison between the human-driven vehicle in the real world and the ADS-under-test in the virtual world, critical scenarios for the ADS-under-test can be revealed. While this approach has the advantage to induce no additional risk for assessing the ADS-under-test, one major drawback of this approach is its open-loop-character, since the evolvement of the real-life scenario, including various trajectories of the incorporated traffic participants (agents) surrounding the virtually engaged ADS-under-test, cannot be stated based on the recorded data. This is caused by the open behavior of the ADS-under-test and the resulting emergent behavior through the interaction of the surrounding agents both among each other and with the ADS-under-test. This applies especially for complex scenarios with various types of traffic participants. For a reliable safety assessment, it is necessary to close the loop. The VAAFO concept presented in [38] represents an approach for closing this open loop. The authors state that based on the concept the advantages of road testing and virtual testing are combined.

According to the authors, VAAFO is observing the environment and compares the behavior of the human-driven vehicle to the virtual behavior of the ADS-under-test. VAAFO consists of four modules, namely the trajectory comparison module, the world correction module, the assessment module and the scenario save module. Within the trajectory comparison module, which can be regarded as a first scenario filter, the desired trajectory of the ADS-under-test and the human-driven vehicle trajectory are compared. If the positional differences between these trajectories over time exceed a certain threshold, the potential critical scenario is forwarded to the world model correction module. Since this position-based filtering can lead to potential critical scenarios, which are not caused by misbehavior of the ADS-under-test, but, e.g., by perception errors of the real vehicle, the world model correction module is necessary to figure out what happened exactly either online or offline [53]. Within the assessment module a further scenario filtering is conducted based on criticality metrics to finally save the remaining scenarios in a database. While VAAFO is based on parallel simulations of the virtual ADS-behavior with limited simulation cycle times to gather all potential critical scenarios, there are cases where the behavior of the ADS-under-test would be falsely regarded as dangerous according to the criticality metrics even in case of consistent perception. In Fig. 3, an exemplary scenario is shown, where the human-driven vehicle behavior and the virtual ADS behavior deviates, while receiving consistent perception information. Since the non-interactive behavior of traffic participant 2 (TP 2) would lead to a virtual collision, the ADS behavior would be falsely classified as dangerous based on the criticality metrics. These scenarios have to be excluded manually in the assessment module.

Another approach for closing the loop utilizing a traffic simulation with short-time behavior models of traffic participants is presented in [39] and illustrated for lane change maneuvers. Using these agent models, the above-mentioned false positive potentially critical scenarios could be avoided. A method for training an algorithm of an ADS-equipped vehicle based on neural networks through comparison of the trajectory of the human-driven vehicle to the virtual desired trajectory of a passive ADS is described in a patent application of Eberle and Hallerbach [54].

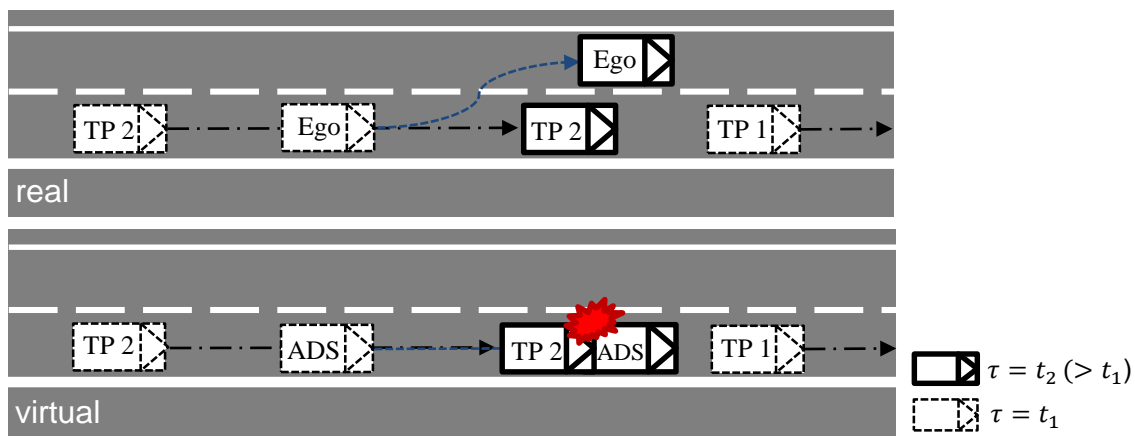

Fig. 3　　　False-positive classification of ADS-behavior as dangerous based on [38]

All these concepts have in common that they solely rely on the specific sensor setup of the human-driven ADS-equipped vehicle and thus influence the ground truth, even after reduction of the included uncertainties. In addition, it can be assumed that the resulting scenario database based on these concepts is only valid for the ADS-under-test passively operated while recording the respective data. The potentially diverging behavior of a different (parametrized) ADS would lead to other scenarios saved in the database. Moreover, to the best knowledge of the authors, there is currently no implementation of this concept for urban environments.

Within this paper, a concept called adaptive replay-to-sim is presented, which has the potential to resolve the ADS-dependence of the aforementioned concepts and could be applied to urban environments. Since the traffic dynamics in urban areas is characterized by the multi-modal interaction of various traffic participants, often ascertainable in spatially bounded areas, trajectory data of all these traffic participants, e.g., recorded by drones, has the potential to serve as ground truth.

Concluding, the idea is to split the execution of a test case (combination of a concrete scenario with evaluation criteria [55]) in two modes, namely a replay-to-sim and an agent-based simulation mode. Within the first mode the ADS-under-test substitutes a vehicle from the real recorded scenario, e.g., recorded by drones or extracted from a stochastic traffic simulation. An observer is tracking the (dis)similarity between the original trajectory of the substituted vehicle and the ADS-under-test. In case the dissimilarity metric exceeds a defined threshold, describing a significant difference between the behavior of the human-driven vehicle and the behavior of the ADS-under-test, an agent-based mode is activated (trigger point). From now on, the behavior and thus the resulting trajectories of the involved traffic participants surrounding the ADS-under-test are determined by the underlying agent models.

This leads to various advantages compared to the present work in the literature, accompanied by some limitations to be solved. Overall, this approach has the potential to allow for revealing the evolvement of the real recorded scenario for the ADS-under-test. In addition, it provides the possibility of an ADS-independent scenario database and thus provides the possibility of a relative comparison between different ADS utilizing a consistent ground truth. Furthermore, it is conceivable to substitute more than one human-driven vehicle with ADS-equipped vehicles to investigate future mixed traffic scenarios due to the additional availability of the vehicle trajectories surrounding the ADS-under-test. In detail, the omniscient view of a scene provided by the trajectory data delivers not only the trajectories of the ADS-equipped vehicle and those of the dynamic objects within the sensor range (accompanied by uncertainties). Rather, in the spatial area of interest it is possible to gather the trajectory information of all dynamic objects relevant to the ADS-driven vehicle, which would otherwise not be accessible. Hence, on the one hand, the origins and destinations of all agents surrounding the ADS-under-test are known, so that agent models can be calibrated for long-term behavior modelling within the respective scenario. On the other hand, this leads to the possibility to not only look in the potential future evolution of the scenario based on the trigger point, but allows for a look in the past for dedicated variations of the initial conditions of all the involved traffic participants due to the existing trajectory data, additionally. This offers the possibility for the exploration of the scenario space around the concrete scenario of interest, which was triggered in advance. In detail, it would be possible to move back in time in the data based on the triggering point upon there is no significant deviation between the human-driven trajectory and the desired trajectory of the ADS to be observed. Now, a parameter variation of the incorporated models within a traffic simulation can be conducted to reflect , e.g., a more aggressive traffic around the ADS-under-test, aiming at revealing the underlying criticality phenomena [22] responsible for the originally observed deviation between human-driver and ADS-under-test. Moreover, this approach can be utilized to figure out, if a false-positive dangerous behavior of the ADS-under-test was tracked or not. In addition, it is conceivable to enable a targeted reinforcement learning process of an ADS using the adversarial characteristics of the surrounding traffic due to the specific parameter variations. Finally, it can be assumed that the ratio between objects relevant for the ADS-under test and objects tracked by the car-specific sensor setup increases with the amount of traffic participants involved in a scenario. Since scenarios in urban areas are expected to be significantly more complex compared to highway scenarios, the advantage of this omniscient view of a scene increases even more.

However, some drawbacks and limitations compared to the above mentioned approaches should be considered. First, the advantage of a consistent ground truth by using trajectory data instead of vehicle data is accompanied by the drawback of not having the (vehicle-specific) world model available and thus necessitates sufficiently valid perception sensor models. Second, inputs for the ADS-under-test are restricted to the recorded variables within the corresponding data set. Moreover, the validity of the incorporated agent models plays a crucial role within this approach. To address these challenges it is conceivable to record both kinds of data in parallel, either in a real traffic environment or in a controlled environment, e.g., on a proving ground. This approach could additionally be utilized for a validation of the purely trajectory data based approach. Moreover, the combined approach could be deployed to enhance the online filtering of potentially critical scenarios to be saved by fusing information of both data sources. For further investigations on this approach, please refer to Section 4.

## 3  Conception

### 3.1  Requirements Specification

Motivated by the two principal research questions tackled in this work (cf. Sec. 1.4), based on which a comprehensive literature research has been conducted (cf. Sec. 2), requirements for the novel simulation-based toolchain for the development and testing of ADS-equipped vehicles in urban environments can be derived. To further encourage a systematic approach for deriving the requirements and, finally, the functional system architecture, an adaption of the abstract human-machine-model [56] is utilized, as shown in Fig. 4.

Since ADS-equipped vehicles are complex systems operating in an even more complex environment, an isolated view on the vehicle is not sufficient any more [56], but rather necessitates a holistic view on the dependencies between the involved entities, namely the ADS-equipped vehicle, the driver/user, the environment and the driving task. With respect to Fig. 4, these dependencies become clear on an abstract level. While the environment influences the ADS-equipped vehicle, e.g., through its topology and weather conditions, the ADS-equipped vehicle moves within the environment, which, in turn, supports the resulting longitudinal and lateral forces due to this movement. Depending on the level of automation the driver or user of the ADS-equipped vehicle interacts with the vehicle over the human-machine-interface and vice versa. Since there is no necessity for a driver/user in case of ADS-engaged vehicles of SAE-Level ≥ 4, the dependencies between the driver/user and the other entities are to be seen as optional and, therefore, represented by dotted lines/arrows. The ADS-equipped vehicle can be divided into the ADS and the residual vehicle (e.g., vehicle dynamics) with interactions amongst each other.

Aiming at a simulation-based toolchain for the development and testing of ADS-equipped vehicles in urban environments, a sufficiently valid virtual representation of different elements of interest constituting the abstract entities shown in Fig. 4 has to be implemented. With respect to the striven capabilities of a hybrid scenario generation and extraction as well as the application of the adaptive replay-to-sim approach various requirements have to be stated. These requirements can be partly categorized by the abstract entities within Fig. 4 and are defined in Tab. 1.

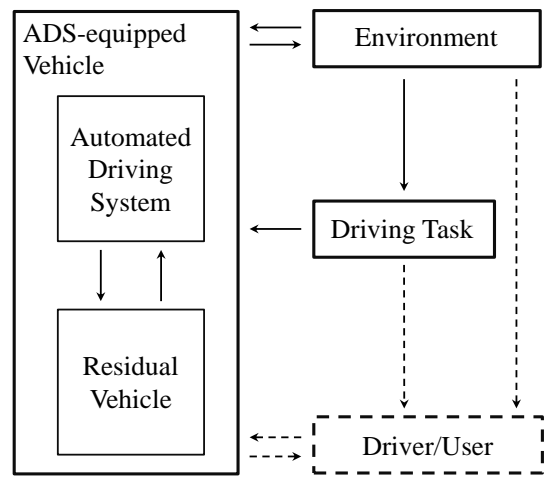

Fig. 4    Abstract human-machine-model (dotted lines/arrows: optional depending on automation level) based on [56]

Within the category of the ego-vehicle representation, two subcategories are defined. First, the toolchain has to provide the possibility to integrate a prototypical ADS [9]. While the main focus is on ADS it is noteworthy that a backward compatibility regarding driving automation systems of SAE-Level 1 and 2 has to be taken into account. Although current validation methods are applicable for a type approval of these systems, concepts and approaches originally conceived for ADS could lead to a significant gain in effectiveness and efficiency within both the development and testing phase of the same. Second, the toolchain has to provide the possibility to integrate a prototypical residual vehicle (vehicle dynamics models, actuator models, etc.).

The environment representation requirements can be divided in two parts. The toolchain has to be able to model both the static and the dynamic part of the environment. Regarding the static part (layer 1–3 of the 6-layer model, cf. Sec. 2.2), the capability to model complex urban traffic spaces is of particular interest. With respect to the dynamic part, requirements regarding the traffic participant models, the so-called traffic agents, surrounding the ADS-under-test, have to be stated (layer 4 of the 6-layer model, cf. Sec. 2.2). To capture the real-world traffic dynamics with sufficient accuracy the traffic agents should be modelled on a submicroscopic level. This means that the behavior and the dynamics model have to be modelled separately to be able to reflect their dependent and independent characteristics [58], [60]. Furthermore, a sufficiently accurate modelling of the multi-modal interaction between various types of traffic agents has to be provided [59]. In addition, the toolchain has to make the programmability of external traffic agent models available [9] to allow for an integration of externally developed, calibrated and validated behavior models. It has to be mentioned that the deployment of a physical-based environment model for, e.g., dedicated perception sensor simulations based on, e.g., material models of objects within the environment, is out of scope at the current state of implementation.

With regard to the simulation control, three different subcategories of requirements are to be mentioned. There has to be a synchronization mechanism, which allows for a communication control between different tools within an instantiation of the toolchain. Additionally, it has to be ensured that the calculation frequency of all incorporated functional modules is below that of the test object (e.g., ADS-under-test) [43] to generate plausible stimuli and, finally, sufficiently valid test case results. With respect to the simulation results processing, further requirements have to be stated. At the one hand, a logging of the quantities of interest has to be ensured while executing a test case. On the other hand, when executing large test case catalogues, an evaluation of the test case results has to be provided in an automated manner. Lastly, the toolchain has to provide the possibility for an export of the generated test case results.

Even though the aforementioned requirements are all contributing to the targeted holistic scenario generation and extraction capability of the toolchain, the scenario engine requirements play a key role regarding its functional deployment. First, there has to be the ability to generate sufficiently valid synthetic data to enrich the scenario database in terms of extent, complexity and impacts of different what-if-scenarios for future mixed traffic. Second, for the deployment of the adaptive replay-to-sim approach (cf. Sec. 2.3) the toolchain has to provide the ability to process real recorded data, especially in form of trajectory data. To be able to derive relevant scenarios this has to be accompanied by a dedicated extraction methodology for urban area scenarios. Finally, to allow for the generation of large amounts of concrete scenarios to be tested, a parametrization of the toolchain must be executable in an automated manner.

The last category is dealing with requirements regarding interfaces, which have to be supported by the toolchain. With a view to data exchangeability, the support of standardized formats and interfaces is of particular interest to ensure a tool-agnostic character of the functional system architecture. Regarding standardized formats especially the OpenDRIVE [61] and the OpenSCENARIO [61] standards have to be mentioned. While the main purpose of OpenDRIVE is the provision of a road network description (layer 1–3 of the 6-layer model, cf. Sec. 2.2), OpenSCENARIO focusses on the description of the dynamic content within a scenario (layer 4 of the 6-layer model, cf. Sec. 2.2). Both standards are developed and maintained by the Association for Standardization of Automation and Measuring Systems (ASAM) [62] within the simulation domain, where further standards with respect to (virtual) validation and verification of ADS-equipped vehicles are developed (e.g., OpenCRG and OpenODD) [61]. Relating to standardized interfaces ASAM OSI [61] and the Functional Mock-up Interface (FMI) [63] standards are of particular interest. First, these standards contribute to a compatibility between an ADS and the variety of driving simulators, ensuring a tool-agnostic character of the toolchain. Second, they support a common interchangeability of implemented simulation models. Finally, aiming at closing the reality-to-sim gap, the toolchain should be applicable across different X-in-the-loop test methods [2], [57].

Tab. 1    Specified requirements for the simulation-based toolchain based on [9], [25], [43], [57], [58], [59]

| Cat. | Subcategory | Definition |
|---|---|---|
| Ego-vehicle Representation | ADS | Possibility to integrate a prototypical Automated Driving System |
| Ego-vehicle Representation | Residual Vehicle | Possibility to integrate a prototypical residual vehicle |
| Environment Representation | Static | Capability to model complex urban traffic spaces |
| Environment Representation | Dynamic (Traffic Participants) | Uninfluenced behavior: Sub-microscopic resolution, Validated behavior models, Capturing of relevant types |
| Environment Representation | Dynamic (Traffic Participants) | Influenced behavior (interaction): Ability of interaction between different types of traffic agents |
| Environment Representation | Dynamic (Traffic Participants) | Integrability: Programmability of external traffic agent models |
| Simulation Control | Scenario Engine | Generation of synthetic data |
| Simulation Control | Scenario Engine | Processing of real recorded data |
| Simulation Control | Scenario Engine | Scenario extraction methodology |
| Simulation Control | Scenario Engine | Automated parametrization of simulation |
| Simulation Control | Simulation Results Processing | Results logging while test case execution |
| Simulation Control | Simulation Results Processing | Automated evaluation of executed test cases |
| Simulation Control | Simulation Results Processing | Simulation results export capability |
| Simulation Control | Synchronization | Calculation frequency below that of test object |
| Interfaces | Data Exchangeability | Support of standardized formats and interfaces |
| Interfaces | Data Exchangeability | Interchangeability of models |
| Interfaces | Test Methods | Usability across different X-in-the-loop test methods |

### 3.2 Derivation of the Functional System Architecture

Based on the stated requirements, a modular, tool-agnostic functional system architecture can be derived, which is shown in Fig. 5. The functional system architecture represents the overall toolchain concept, striving for an approach supporting both the development and testing phase of ADS-equipped vehicles, especially in urban areas. Instantiations of various modules of the functional system architecture for gaining first results with respect to the two focused research questions are described in Section 4.

The functional system architecture is split into three major parts, namely the ego-vehicle representation, the environment representation and the simulation control. Since the sufficiently valid representation of the traffic space and dynamics surrounding the ADS-under-test (layer 1–4 of the 6-layer model, cf. Sec. 2.2), characterized by the multi-modal interaction of various types of traffic participants, is a central building block for the scenario generation and the subsequent scenario extraction, a traffic simulation constitutes the part of the environment representation. This part can be further divided into two functional modules containing both the static and the dynamic part of the environment representation. The static part (traffic space) can be implemented, e.g., by using an OpenDRIVE file [61] or a Lanelet2 map [64]. This static description serves as input for the dynamic part of the environment representation. The dynamic part, the so-called traffic flow model, can be further subdivided into the traffic flow calibration module and the traffic agent models module. The combination of these two modules aims at modelling the multi-modal interaction of traffic participants surrounding the ADS-under-test with sufficient accuracy. Whilst the traffic flow calibration module is characterized by the parametrization of various macroscopic [59] quantities of interest, the traffic agent models module is included to reflect the behavior and dynamics of the various traffic participants of interest on a (sub-)microscopic level. The necessary information for the parametrization of the macroscopic quantities of interest within the routing, compositions and volumes module is delivered by data available in the real recorded data module. This data serves also as input for the calibration of the different traffic agent models in form of pedestrian models, bicyclist models and models representing different vehicle types. At the current state of implementation, the integration of a public transportation model (e.g., trams) can be seen as optional, since our first investigations are focusing on unsignalized intersections without this mode of transport. Busses can in turn be modelled using a customized parametrization of the vehicle models. Besides the distinction of the vehicular traffic by type in form of cars, trucks and busses, it is possible to distinguish the vehicular traffic by automation level. For an insertion of vehicles equipped with driving automation systems or ADS within the simulation there are two possibilities: for a phenomenological modelling, an adjusted calibration of the vehicular models, originally developed to reflect human-driven vehicles, can be conducted. For a more sophisticated modelling, multiple ego-vehicle instances with individual parametrizations can be deployed within the simulation.

The modelling of one or more ego-vehicle representations interacting with the environment representation can generally be divided into the functional module representing the ADS and the residual vehicle module. It is noteworthy that in case of investigations focusing on the human-machine-interface (HMI) between driving automation systems and the respective driver, the ego-vehicle representation has to be extended by an adequate driver model. In case of investigations concerning ADS-engaged vehicles, the ADS-under-test is located in the ADS-module further divisible in the localization module, the perception module, the planning module and the motion control module. The ADS-under-test is interacting with the residual vehicle through commands from the motion control to the actuation models, which feed into the vehicle dynamics model. The vehicle dynamics model in turn serves as input for the ADS-under-test and residual vehicle functions of interest (e.g., HMI or light), which are optionally fed by the ADS, additionally.

In terms of a holistic scenario generation and extraction enabled through the functional system architecture, the simulation control plays a key role. While the interplay of the ego-vehicle simulation and traffic simulation aims at a sufficiently valid execution of the defined test cases, the simulation control cares about the question, which are the relevant scenarios to be tested, defines the results processing approach and orchestrates the ensemble of the various functional modules. Within the scenario engine module, the real recorded data is transformed to a specific form, able to serve as feasible input for different modules. At the current state of implementation, the real recorded data is limited to trajectory data of unsignalized intersections, but can be extended to other data sources in future. This data serves as input for both the traffic flow calibration module and the traffic agent models module. Furthermore, it feeds into the scenario extraction module, whose current state of implementation will be introduced in Section 4.2, briefly.

Utilizing the simulation results of the coupled ego-vehicle and traffic simulation, the synthetic data module conditions the simulation data in a way that allows for a subsequent processing by the scenario extraction module in order to obtain a hybrid scenario data base further described in Section 4. The output of the scenario extraction module serves as input for the simulation parametrization and execution module. There is the possibility to either define a concrete test case to be executed within the coupled simulation (e.g., specified via an OpenSCENARIO file) under use of the results processing module or to use the real recorded data for an adaptive replay-to-sim (for implementation considerations cf. Section 4). Moreover, the simulation parametrization module communicates with the results processing module to combine a concrete scenario with evaluation criteria to derive a concrete test case. Inside the results processing module the data logging module receives the quantities of interest to be logged and evaluated either online during test case execution or offline. The results export module controls a valid storage of the recorded results in the specified repository. Finally, the synchronization module has the task to ensure a proper data exchange during the execution of a test case, e.g., with different solvers based on inputs of the simulation parametrization and execution module (e.g., test case specification [55]).

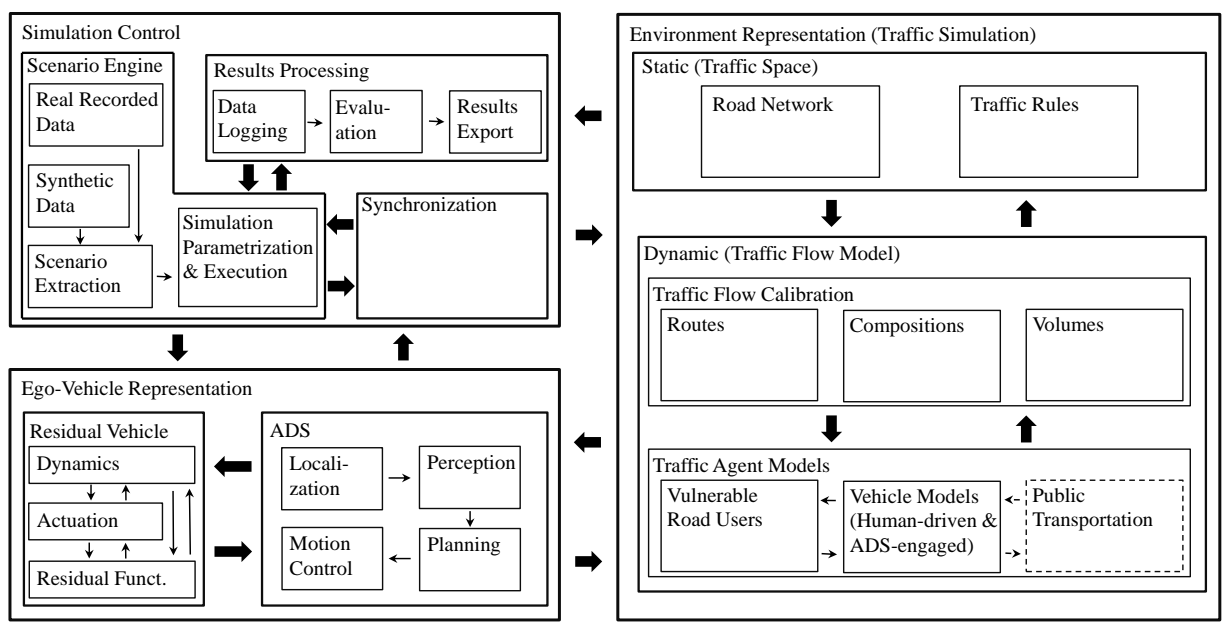

Fig. 5    Functional system architecture (dotted lines/arrows (functional dependences): optional modules)

## 4　Results

This chapter gives an overview about the current state of implementation regarding various modules of the functional system architecture derived in Section 3.2. The explanations of the implementation and the description of first results are conducted with reference to both respective relevant requirements and contributions related to the stated research questions.

Since the simulation control module plays a key role regarding a functional deployment of a holistic scenario generation and extraction capability of the toolchain (cf. Sec. 3), first implementations are focusing on an instantiation of this module. Thereby, the fulfillment of the requirements concerning the processing of real recorded data in form of trajectory data, a deployment of a dedicated extraction methodology for urban area scenarios and the generation of sufficiently valid synthetic data are of particular interest for both stated research questions. Additionally, due to the interactions between the environment representation module and the scenario engine module, investigations concerning the deployment of a sufficiently valid traffic simulation are important from the very beginning, as they can be seen as prerequisite for both the synthetic data generation and the implementation of the adaptive replay-to-sim approach.

### 4.1　Real Recorded Data Module

Taking a deeper look on the fulfillment of the requirement regarding the processing of real recorded data, the current state of implementation with respect to the real recorded data module and the scenario extraction module is particularly relevant. Due to the demand for real recorded data in form of trajectory data for the investigation on both stated research questions, a suitable dataset has to be selected in a first step. As result, the intersection Drone dataset (inD dataset) [26] was chosen due to various reasons. The main advantages of the inD dataset compared to other similar datasets are its size, representativeness and accuracy [26]. The inD dataset consists of trajectories of more than 11500 traffic participants including cars, trucks and busses as well as about 5000 pedestrian and bicyclist trajectories recorded at four different intersections in Aachen, Germany. A scene within the Bendplatz traffic space including labeled traffic participant trajectories is shown in Fig. 6, exemplarily.

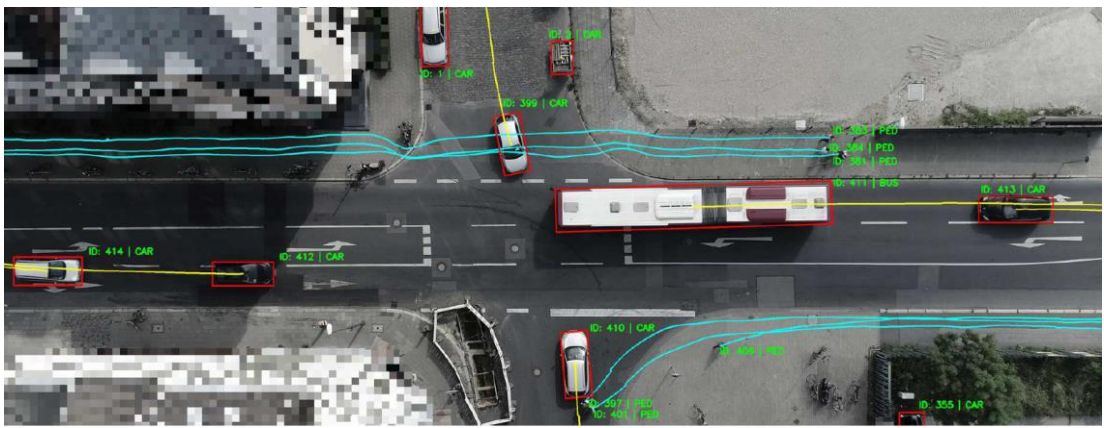

Fig. 6　　Birds-eye view of Bendplatz traffic space (lat.: 50.78206, long.: 6.07120) with labeled TP trajectories [26]

Utilizing camera-equipped drones for recording the traffic, naturalistic behavior can be observed, as the traffic participants do not notice measurements taking place [26]. The availability of this dataset leads to the possibility to serve as feasible input for different modules, namely the scenario extraction module and the traffic simulation module.

### 4.2 Scenario Extraction Module

With respect to the scenario extraction module, a methodology has been developed able to transfer the multivariate time series data constituting the inD dataset to complex relevant concrete scenarios for subsequent testing of ADS-equipped vehicles, as shown in Fig. 7. Thereby – compared to previous work – the developed approach allows for extracting concrete scenarios, capturing the multi-modal interaction of various traffic participants, in form of vehicle-to-vehicle (v2v), vehicle-to-pedestrian (v2p) and vehicle-to-bicycle (v2b) scenarios with an allocation to a set of functional scenarios.

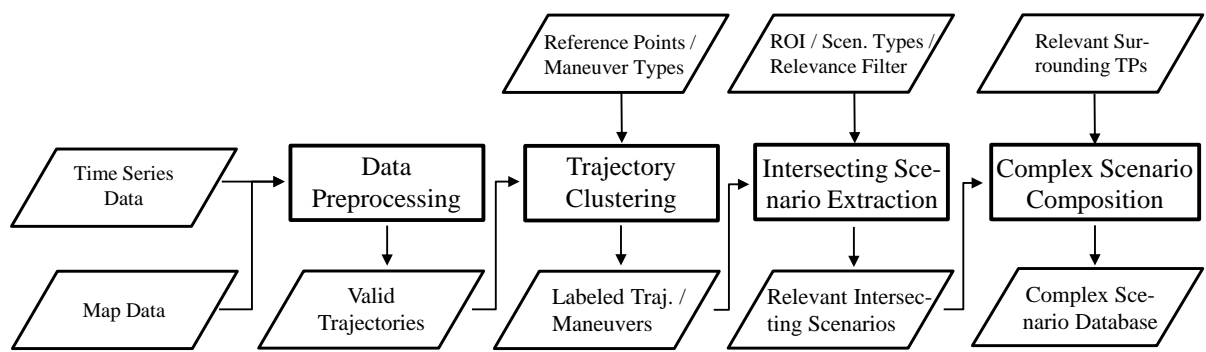

Fig. 7     Overall approach for the extraction of relevant urban area scenarios

In the first step of the scenario extraction module, a preprocessing is implemented to remove incorrect data, e.g. outliers, within the inD dataset. This filtering process is based on the TP class (car, truck, bus, pedestrian, bicycle), the velocity data and the total lifetime in frames of the corresponding traffic participant within the recording. An example for such an outlier is the pedestrian with ID 60 in record 17, who suddenly occurs nearby the intersection center with a high velocity and disappears after around 2 seconds, resulting in a very implausible behavior. This kind of trajectories might be caused by, e.g., tracking errors of the computer vision algorithms. In such cases, the trajectories are classified as invalid and filtered out for the subsequent extraction steps. However, it has to be mentioned, that this rule-based filtering could theoretically lead to both undetected relevant scenarios or forwarding irrelevant scenarios to the subsequent extraction steps.

The valid trajectories are forwarded to the trajectory clustering module where an allocation of single trajectories to predefined maneuver types is performed based on reference points for the traffic space under investigation. Depending on the TP class both the underlying reference point set and the possible maneuver types differ. Thereby, the labeling of the trajectories is determined by the respective compass directions of passed intersection branches. At the current state of implementation, the respective reference point set has to be chosen manually per traffic space. These reference points subsequently serve as fixed centroids within a k-means based clustering process to allow for a robust allocation of single trajectories to defined maneuver types. In Fig. 8, the resulting vehicle trajectory labels and corresponding identified maneuver types are shown for record 16 within the Bendplatz traffic space, exemplarily. The trajectory labels "EE" and "SS" are caused by two not assignable trajectories. Thereby, the trajectory labeled with "SS" and subsequently assigned to the maneuver type "special" is constructed by a theoretically illegal U-Turn of a vehicle nearby the intersection center and can be seen as corner case. It is of note, that for this recording only two of 270 trajectories could not be filtered out or assigned to a specified maneuver type in an automated manner. Moreover, the question arises whether it is generally feasible to define separate maneuver types for such corner cases, since they are rare and their relevance determination often necessitates the inclusion of expert knowledge.

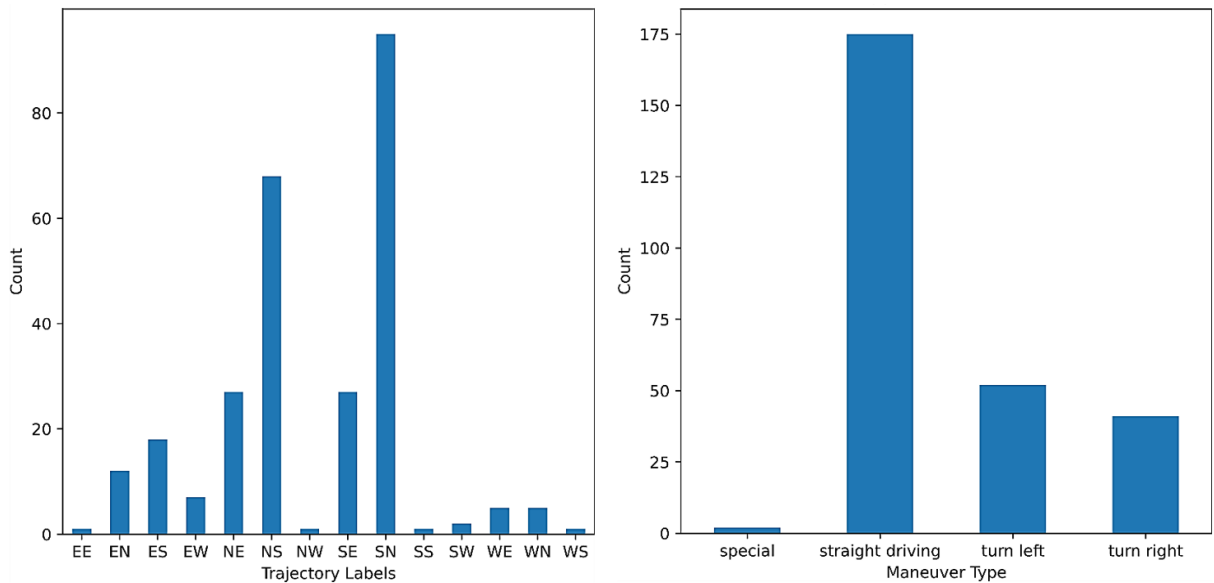

Fig. 8     Vehicle trajectory labels (left) and maneuver types (right) (record 16)

The entirety of labeled valid trajectories with respective maneuver types of the different TP classes serves as input for the intersecting scenario extraction module. Before introducing the sub-steps of this module, it has to be mentioned that in urban areas the safe operation of ADS-equipped vehicles within multi-modal scenarios including, e.g., pedestrians and/or bicyclists, will be a decisive aspect with respect to the safety proof. Due to this, the implementation of the intersecting scenario extraction module is conducted in a way that allows for an extraction of v2v, v2p and v2b scenarios, referred to as scenario categories. Since every vehicle of the TP class car in the dataset can be seen as potential ego-vehicle, a moving region of interest (ROI) in form of a circle with a radius of 15 meters is defined around it while moving through the intersection to capture possibly relevant challenging traffic participants in the surrounding. Thereby, intersecting scenarios are defined as scenarios including two traffic participants with intersecting trajectories, namely a vehicle and a surrounding traffic participant. First, based on the TP class combination of the respective scenario, a mapping to the aforementioned scenario categories (v2v, v2p, v2b) is done. Second, based on the derived scenario category a further allocation is conducted to a corresponding functional scenario type. At the current state of implementation there is a set of 10 functional scenario types defined for the v2v scenario category based on [27] and [65], see Fig. 9. These scenario types apply as well for the v2b scenario allocation. Additionally, for the v2b scenario category a further scenario type called "not-cross" is added to capture those scenarios within which the corresponding bicyclist just rides on the sidewalk and thus can be assumed as less relevant in terms of safety validation of the ADS-under-test. Concerning v2p scenarios, at the current stage, there is a binary classification of the related concrete scenarios into the functional scenario types "cross" or "not cross" depending on the existence of an intersection point between the respective vehicle and pedestrian trajectory.

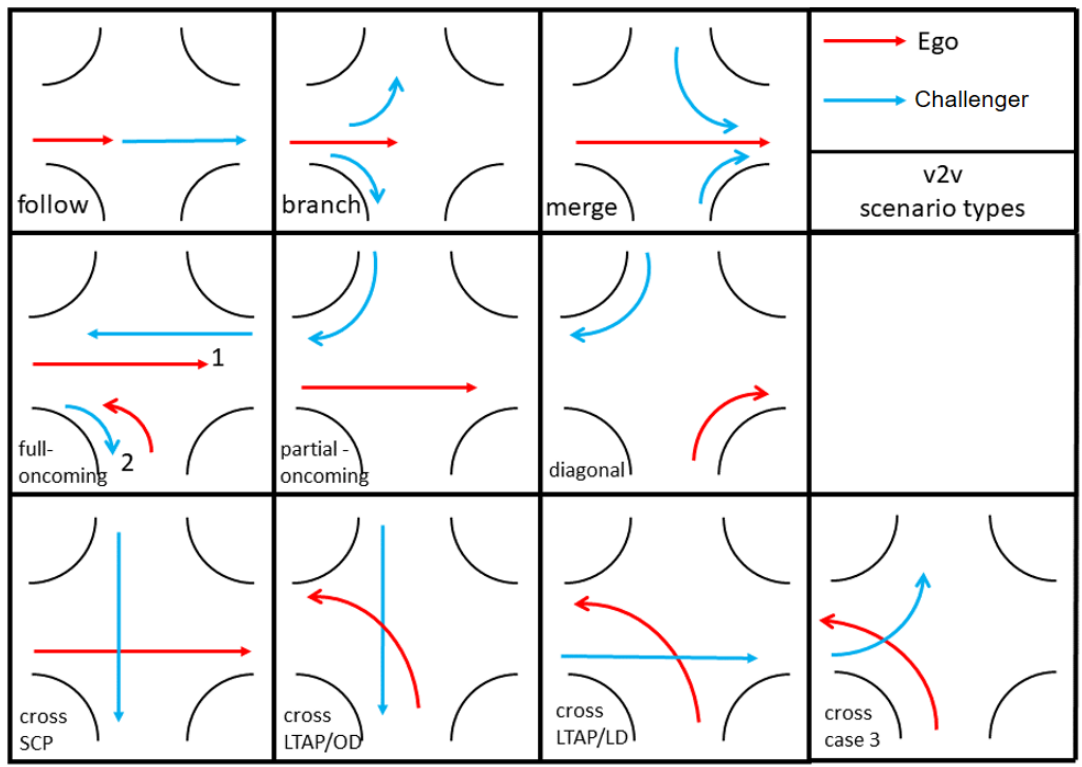

Fig. 9    Functional scenario types (SCP: straight crossing path; LTAP: left turn across path; OD / LD: oncoming- / lateral direction) based on [27], [65]

Aiming at reducing the amount of data while increasing the amount of knowledge about the same, the entirety of intersecting scenarios is filtered by their relevance through evaluating the Post-Encroachment Time (PET) [69] of the included traffic participants. Based on the relevant frame indexes of the included traffic participants in the corresponding intersecting scenario, the scenario length is tailored, so that a shorter scenario segment with higher information density remains. In Fig. 10, an exemplary PET distribution within the different vehicle-to-bicycle scenario types is shown for record 16 using box-plots. It can be observed that there is a high scatter of the PET values over the different scenario types. Additionally, the absolute scatter of the respective functional scenario is varying significantly depending on the type. At the current state of implementation, a PET threshold of 6.5 seconds is implemented for subsequent relevance filtering of the intersecting scenarios according to [66], aiming at a compromise between the amount of scenarios to be tested and not neglecting potentially critical scenarios. The authors are aware that for purely human-operated scenarios PET threshold values ≤ 1.5 s are considered as potentially critical, typically [67], [68], [69], but a more conservative assumption is reasonable due to the open behavior of the ADS-under-test.

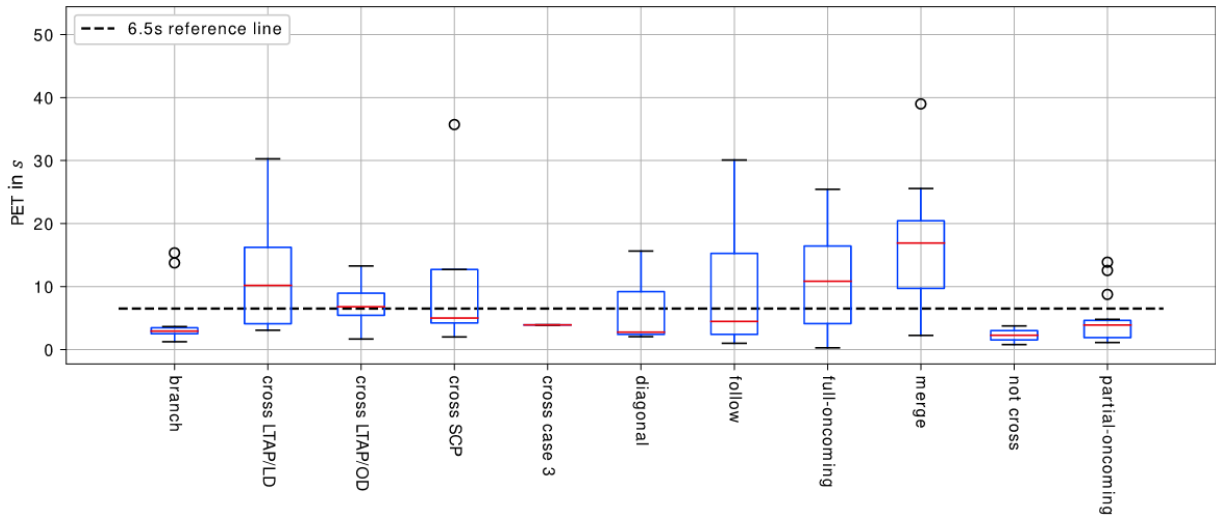

Fig. 10    PET distribution of v2b scenario types (record 16)

The tailored intersecting scenarios are the starting point for the last sub-step within the scenario extraction module, namely the complex scenario composition. Within this step, the moving ROI is again applied for all vehicles of the relevant intersecting scenarios to capture the surrounding traffic participants over the duration of the corresponding scenario. This leads to multi-modal concrete scenarios of different complexity allocated by the introduced functional scenario types, which are handed over to the complex scenario database as OpenSCENARIO files [61]. Through the availability of clustered scenarios it is possible to derive logical scenarios within each functional scenario, e.g., in form of probability distributions. This allows, e.g., for a probability-based sampling of the logical scenarios to derive new concrete scenarios not captured by the dataset for exhaustive testing. Concerning the stated requirements for the target toolchain, this approach contributes to the capability of the toolchain to process real recorded data and extract urban area scenarios for subsequent testing of ADS-equipped vehicles. Moreover, the extracted scenarios can serve as input for the adaptive replay-to-sim approach.

### 4.3    Synthetic Data and Traffic Simulation Module

To allow for a synthetic data generation by the toolchain, a deeper look has to be taken on the possibilities to deploy a traffic simulation capable of capturing the traffic dynamics within urban areas in a sufficiently valid manner. The implementation of the traffic simulation module within the overall toolchain plays a key role for the validity of the generated synthetic data. Moreover, the included traffic agent models are of particular importance for an implementation of the adaptive replay-to-sim approach.

With respect to available tools, microscopic traffic simulators have the potential to fulfill the stated requirements regarding a sufficiently valid environment representation (cf. Sec. 3.1). Furthermore, these tools are already used for investigations within simula-tion frameworks for developing and testing of ADS-equipped vehicles focusing on highway environments (cf. Sec. 2.1). Compared to established simulation frameworks, which cover vehicle testing issues and focus on vehicle, sensor and driver models, microscopic traffic simulators provide the possibility to generate large-scale urban traffic with multi-modal interactions of different types of traffic participants [9], [70]. Some out of the variety of these microscopic traffic simulators [70], [71] – originally developed for, e.g., traffic infrastructure planning and traffic flow optimization [72] – are pushing into the field of development and testing of ADS-equipped vehicles [60]. In this context, PTV Vissim (Vissim) [73], Eclipse SUMO (SUMO) [74] and Aimsun [75] can be named. The tool providers mention the capability of simulating automated and connected vehicles and contribute to research projects in the field of ADS development and testing [76], [77]. Therefore, we are working on an in-depth evaluation of microscopic traffic simulators in the context of developing and testing ADS-equipped vehicles in urban areas to be published in the near future.

Based on literature research and first implementations within Vissim and SUMO, an analysis of the potential of these tools has been conducted. With respect to the stated research questions, the capability of the traffic simulation to reflect the traffic dynamics on a macroscopic level, e.g., for a complete intersection traffic space, is of particular importance to serve as synthetic data source. If the traffic dynamics can be reflected in a sufficiently valid manner, it is conceivable to execute long-term simulations to increase the representativeness of a scenario database. In addition, it would be possible to extrapolate the traffic dynamics in a more complex direction through parameter variations for, e.g., generating scenarios with a more aggressive TP behavior. Moreover, there would be the possibility to substitute human-driven vehicles by ADS-driven agents to extend a scenario database in terms of future mixed-traffic scenarios.

After conducting first investigations, the results can be described as follows. With respect to the striven capability of modelling complex urban traffic spaces it has to be distinguished between the general possibilities provided by the tools and compatibility with standardized formats. Both Vissim and SUMO allow for a flexible generation of road networks within the respective user interface. While the road network of Vissim is based on links and connectors, it is based on nodes and edges in case of SUMO. Regarding the modelling of urban intersections this leads to the possibility of modelling smooth routes for the vehicles to follow in Vissim, while the ones are angular when modelled using SUMO, as shown in Fig. 11. Both Vissim and SUMO allow for an import of OpenDRIVE files with some limitations, e.g., regarding the representation of road markings and variable lane widths.

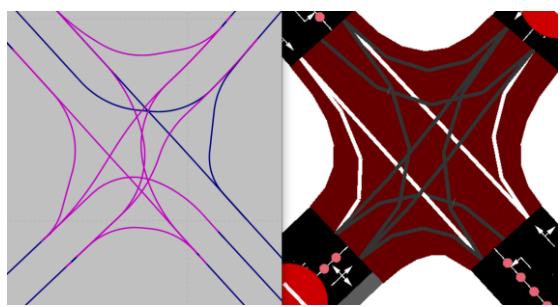

Fig. 11    Modelling of the Bendplatz traffic space with Vissim (left) and SUMO (right)

The road network modelling capability has a direct impact on the dynamic part of the environment representation and thus the fulfillment of the respective requirements, since the trajectories to follow by the vehicular traffic are defined by the road network. While SUMO offers the possibility to implement various car-following models for the modelling of driving in a stream on a single lane [78], Vissim utilizes the psycho-physical Wiedemann model [79]. Vissim and SUMO offer different possibilities to model the lane change behavior of vehicles. In addition, the lateral in-lane behavior can be manipulated to some extent, e.g., for modelling the overtaking of a bicycle by a car within a defined lane. Regarding the modelling of pedestrians, SUMO offers two different models, namely the "nonInteracting" model and the "striping" model. With respect to Vissim, it is possible to either model pedestrians based on an adapted parametrization of the Wiedemann model or to use the Viswalk module. Within Viswalk, the movement of a pedestrian is defined by a Social-Force model [80], which allows pedestrians to walk to their destination without the network model predefining their trajectories. The basic concept of this model is the determination of motion of pedestrians by so-called social forces. In this context, social forces describe the motivation of pedestrians to move in a specific direction based on their destination and other static and dynamic objects in their surroundings. Thereby, pedestrians are exposed to different attractive and repulsive forces whose sum determines the respective movement. For the modelling of bicyclists, neither of both simulators offers a dedicated model. Consequently, an adapted parametrization of car-following models has to be implemented [78], [79].

In the left part of Fig. 12, the vehicular trajectories of recording seven at Bendplatz traffic space are shown. On this qualitative level, it becomes clear that – compared to highway environments – a rule-based behavior still seems to predominate the tactical level of the respective drivers. However, a significant variation regarding the trajectories of drivers following the same route on the same lane can be observed, what is in accordance with [8]. This raises the question whether the car-following models available in Vissim and SUMO are able to reflect the urban (vehicular) traffic dynamics in a sufficiently valid manner, since the (non-varying) routes of the network model predefine the corresponding trajectories. At the current stage of implementation, we are still conducting in-depth investigations regarding this question and possible consequences related to the stated requirements. On the one hand, it has to be analyzed what sufficiently valid means in terms of synthetic data generation for the development and testing of ADS-equipped vehicles in urban areas utilizing traffic simulations (macroscopic view). On the other hand, it has to be figured out how feasible the available agent models are for conducting an adaptive replay-to-sim of concrete scenarios (microscopic view). It is noteworthy that the pedestrian and bicyclist trajectories are varying even more and separate investigations have to be conducted for the respective underlying agent models.

Based on qualitative observations and quantitative analysis of the dataset, we saw potentials in a more flexible and variable modelling of the vehicular traffic in such urban traffic spaces. Hence, we decided to develop a Social-Force model for urban vehicular traffic based on [7] and [81]. The developed extensions focus on the capability to model vehicular traffic at unsignalized intersections. Therefore, significant adaptions to the model were implemented to allow for the consideration of priority rules based on additional social forces. Moreover, relevant adaptions regarding the social forces determining the velocity within curves were made. Furthermore, significant adaptions regarding the interaction of vehicles surrounded by pedestrians and bicyclists were implemented. Finally, relevant changes were applied to the lateral in-lane behavior of the corresponding vehicle. Additionally, the Social-Force part of the model was coupled with a kinematic single-track model to allow for a more realistic movement of a vehicle in space. Subsequently, the parametrization of the model was conducted based on the trajectory data of Bendplatz traffic space using a genetic algorithm. With respect to Fig. 12, first results look promising, since the traffic participants controlled by the model (right) qualitatively capture the variance of the vehicular traffic quite well compared to the real recording (left). This implies a good approach for modelling the multi-modal interactions and the consideration of priority rules as well, since the real pedestrian and bicyclist trajectories were included within the parameter optimization process. It has to be mentioned, that the results are to be seen as a proof of concept and an in depth-investigation of this model is to be conducted.

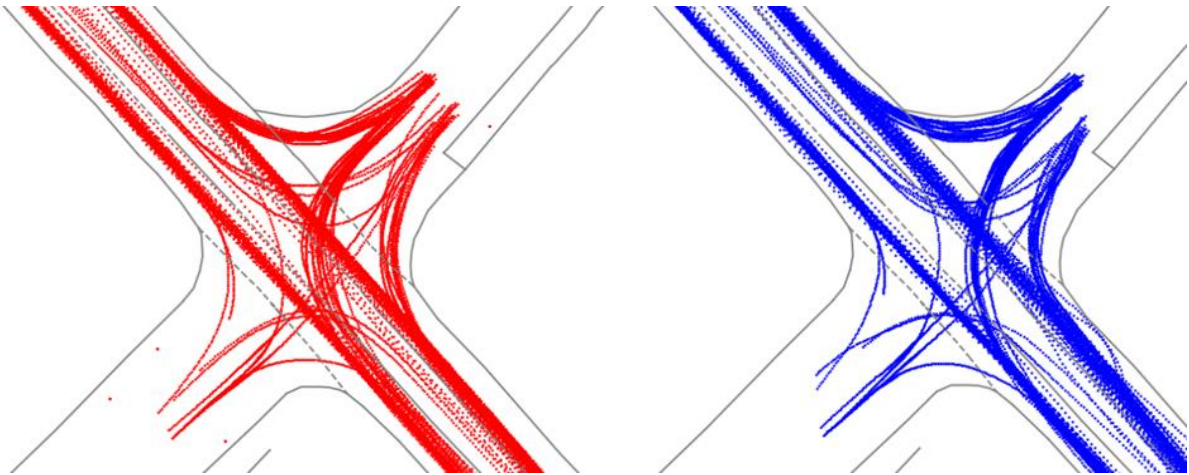

Fig. 12    Real recorded vehicular trajectories (left) and simulated trajectories (right) at Bendplatz traffic space (recording 7)

The aforementioned investigations contribute especially to the answer of the first research question, i.e., the synthetic data generation by the toolchain. To gain knowledge regarding the answer of the second research question concerning the adaptive replay-to-sim approach, first implementations have been conducted as well using Vissim and SUMO. In detail, we analyzed how realistic specific concrete scenarios of the inD dataset can be approximated by simulations of Vissim and SUMO in an agent-based manner. To gain dedicated knowledge about the fulfillment of the requirements regarding the uninfluenced and influenced behavior of the traffic participants (cf. Sec. 3.1), six representative scenarios including multi-modal interactions of different traffic participants were selected. Thereby, the first three scenarios capture the uninfluenced behavior of vehicles, pedestrians and bicyclists, while the last three scenarios capture the vehicle-to-vehicle, vehicle-to-pedestrian and vehicle-to-bicycle interaction.

In Fig. 13, an illustration of the investigated concrete vehicle-to-pedestrian scenario at the Frankenburg traffic space is shown. Within this scenario, a pedestrian initially walks over the sidewalk on the upper side of the eastern branch and then moves over the pedestrian crossing towards the lower side of the eastern branch. Consequently, a car driving on the eastern branch towards the pedestrian crossing has to brake for the pedestrian and thus an interaction occurs. After the pedestrian has left the conflict area in front of the car, it accelerates and makes a right turn towards the northern branch. This scenario is chosen for the following explanations, since it includes both an uninfluenced and an influenced behavior part of the two dominating traffic participant types within the recording.

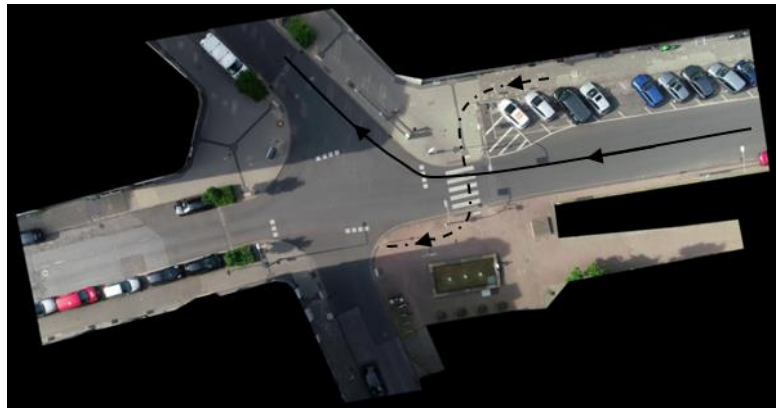

Fig. 13　Illustration of vehicle-to-pedestrian scenario at Frankenburg traffic space (lat.: 50.76835, long.: 6.10228; pedestrian: chain dotted line, car: solid line) based on [26]

In Fig. 14, the trajectories of the real recorded traffic participants as well as the trajectories simulated by Vissim and SUMO occurring for this scenario are shown. To allow for a meaningful comparison, the Wiedemann model was used in both simulators to approximate the real vehicle behavior. Concerning the modelling of pedestrians, the Social-Force model was deployed in Vissim while in case of SUMO the Model "striping" was utilized. Based on Fig. 14, first conclusions can be drawn. First, it can be observed that the simulated trajectories of both Vissim and SUMO are in good accordance with the real trajectory as long as the vehicle is approaching the pedestrian crossing and drives in the middle of the lane. Afterwards, the deviations between the real and simulated vehicle trajectories increase, since the real vehicle does not drive exactly in the middle of the lane. These deviations in the spatial dimension are caused by the network model. The greater deviation and the more angular trajectory of the vehicle simulated by SUMO is caused by the connecting lanes within the node. In case of the pedestrian trajectories, this also applies for SUMO, while regarding Vissim there is no predefinition of the pedestrian trajectories through the network model and thus a good approximation of the real trajectory can be observed.

With a view to the real and simulated acceleration courses of the vehicle in Fig. 15, it becomes clear, that Vissim allows for a parametrization, which seems to approximate the driving behavior in a feasible manner over the scenario evolvement. Since the Wiedemann model is applied in both simulators, the greater deviation of SUMO might be caused mainly by the network modelling, available acceleration states of the Wiedemann model and possibilities to influence the longitudinal dynamics of the vehicle beyond the car-following model.

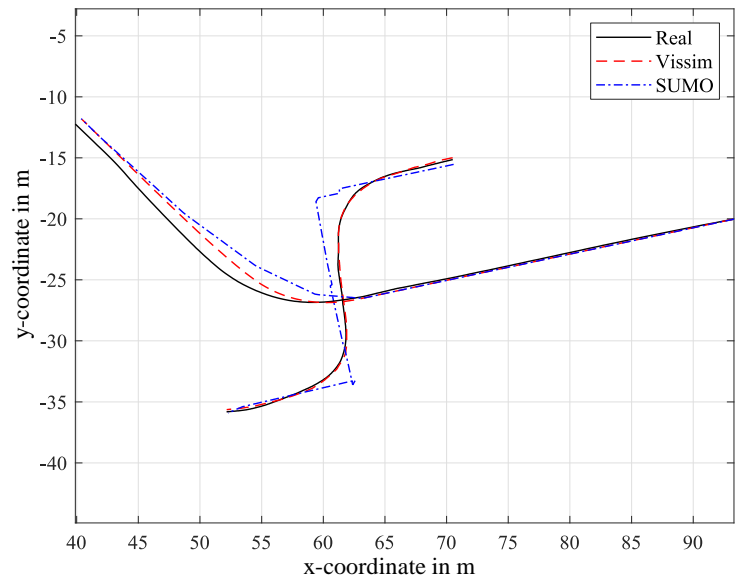

Fig. 14    Trajectories of real and simulated traffic participants

While Vissim allows for the manipulation of the acceleration, e.g., through reduced speed areas, there is no comparable option in SUMO. Moreover, the less sophisticated interaction modelling between vehicles and pedestrians results in acceleration steps and even an emergency braking in case of SUMO. Furthermore, the greater deviations from the real recorded trajectory are caused by the availability of only four discrete acceleration values, which the car-following model can be assigned to within SUMO.

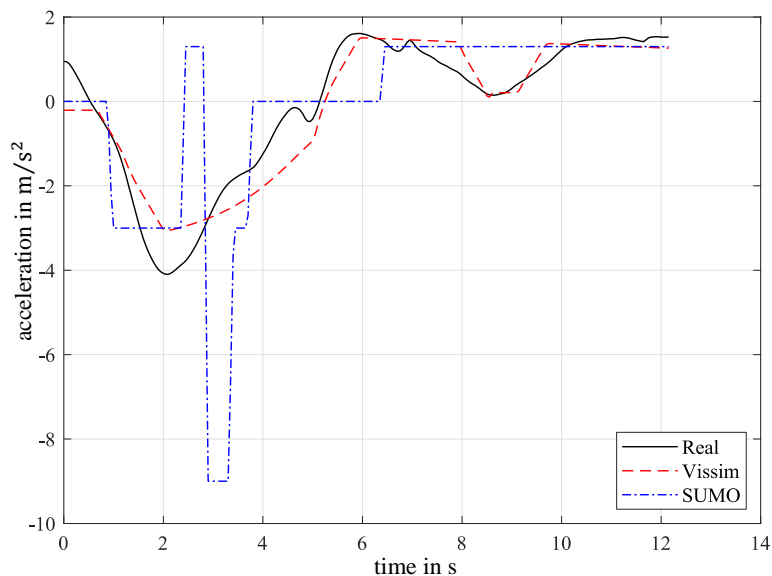

Fig. 15    Acceleration courses of real and simulated vehicles

Based on the current stage of implementation it can be summed up that available microscopic traffic simulators offer the potential to support the development and testing of ADS-equipped vehicles in urban environments. While the approximation of real traffic participants behavior looks promising for specific concrete scenarios especially in case of an agent-based simulation by Vissim, further research has to be conducted on this use case. Particularly, the observed variance of vehicle trajectories within the dataset cannot be represented sufficiently within established traffic simulators so far. Thereby, first results of the developed Social-Force model for urban vehicular traffic look promising for tackling this problem. Since the capabilities of the available traffic simulators can be seen as good basis in the context of both stated research questions, future work will focus on further closing of the gap between stated requirements for the toolchain and available capabilities by the traffic simulators. As next step, the developed Social-Force model will be integrated within one of the examined traffic simulators striving for the possibility of generating sufficiently valid synthetic data and implementing the adaptive replay-to-sim approach with an ADS-equipped vehicle representation.

## 5   Conclusion and Outlook

Within the scope of this paper, the overall concept of a novel simulation-based toolchain for the development and testing of ADS-equipped vehicles in urban environments was presented. Two research questions aiming at a holistic scenario generation and extraction methodology by the toolchain were stated, which defined the scope of investigations within this paper. The first question puts emphasis on the ability of the toolchain to serve as synthetic data source within the development phase of ADS-equipped vehicles to enrich a scenario database in terms of extent, complexity and impacts of different what-if scenarios for future mixed traffic. The focus of the second research question is on the implementation of a so-called adaptive replay-to-sim approach. This approach aims at combining the individual advantages of real recorded data and an agent-based simulation.

Based on the stated research questions and a comprehensive literature review, requirements for the toolchain were derived systematically, inspired by an adapted abstract human-machine-model. As a result, 17 top-level requirements could be stated within nine subcategories and four main categories, respectively. This requirement specification formed the basis for the subsequent derivation of a modular, tool-agnostic functional system architecture representing the overall toolchain concept. Having the functional system architecture at hand, instantiations of various functional modules of particular importance for answering the stated research questions could be implemented. This led to first results regarding the real recorded data module, the scenario extraction module as well as the synthetic data and traffic simulation module.

With respect to the real recorded data and scenario extraction module it could be demonstrated that naturalistic driving data in form of traffic participant trajectories can contribute as important data source within an extraction methodology for ADS-relevant urban area scenarios. Thereby, the current state of the developed methodology allows for an extraction of vehicle-to-vehicle, vehicle-to-pedestrian and vehicle-to-bicycle scenarios and a subsequent assignment to a functional scenario set. This allows, e.g., for a creation of logical scenarios and a subsequent probability-based sampling to derive new concrete scenarios not captured by the dataset for exhaustive testing. Furthermore, the extracted relevant concrete scenarios can serve as input for the adaptive replay-to-sim implementation.

Relating to the synthetic data and traffic simulation module, first investigations focused on answering the question whether available microscopic traffic simulators have the potential to serve as basis for both a synthetic data source and the implementation of the adaptive replay-to-sim approach. It could be demonstrated that the examined traffic simulators cannot cope with the increased variance of vehicle trajectories within urban traffic spaces compared to highway environments. Therefore, the concept and first implementations of a Social-Force model for urban vehicular traffic was presented, showing promising results for tackling the aforementioned problem of variable vehicular traffic. Whilst these investigations are particularly relevant for a sufficiently valid synthetic data generation by the traffic simulation further analyses regarding the adaptive replay-to-sim approach were conducted. Therefore, the capability of the microscopic traffic simulators PTV Vissim and Eclipse SUMO to simulate specific concrete scenarios in an agent-based manner was examined. The scenarios focused on the uninfluenced and interactive behavior of different types of traffic participants, whereby a thorough analysis was given for a vehicle-to-pedestrian scenario. As result, the general potential to execute concrete scenarios by the traffic simulators could be demonstrated, whereby Vissim outperformed SUMO in terms of approximating the reality. Possible reasons for the remaining deviations were revealed and can serve as starting point for extensions of the simulators.

Based on the derived results, future research directions have been identified. First, the implementation of the scenario extraction module shall be extended by advanced clustering techniques and a more sophisticated relevance filter. Second, the derivation of logical scenarios within each functional scenario, e.g., in form of probability distributions is of particular interest. This would allow, e.g., for a probability-based sampling of the logical scenarios to derive new concrete scenarios not captured by the dataset for exhaustive testing. Moreover, based on the promising results concerning the traffic simulation module, the next step is the enabling of the scenario extraction methodology for processing synthetic data generated by the traffic simulation module. A deeper look should be taken on further data sources of real recorded data, since the available trajectory data has advantages concerning the representativeness, but limitations concerning the occurrence of corner cases as long as the amount of data itself is limited.

Striving for a sufficiently valid synthetic data generation, an in-depth investigation of established microscopic traffic simulators in the context of developing and testing ADS-equipped vehicles in urban environments has to be conducted based on the derived results. Thereby, the macroscopic dimension in terms of approximating macroscopic traffic metrics like traffic flow or traffic density plays a crucial role for the ability to serve as synthetic data source and enrich the scenario database in terms of extent, complexity and impacts of what-if scenarios for future mixed traffic. The microscopic dimension including the execution of concrete scenarios in an agent-based manner is of particular interest for the adaptive replay-to-sim approach. The decision to be taken concerning sufficient validity of the traffic simulation in the context of developing and testing ADS-equipped vehicles in urban areas rises another research question. Contributing to the answer of this question the developed Social-Force model is to be integrated in one of the established traffic simulators to close the gap between current capabilities and stated requirements for the toolchain. Finally, one key part of the functional system architecture yet not instantiated, namely the ego-vehicle representation, will be implemented in the near future. This will allow for both a more sophisticated synthetic data generation in terms of future mixed traffic and the completion of the adaptive replay-to-sim implementation.

## 6  Acknowledgement

We want to thank Dr. Ulrich Eberle and Dr. Stefan Berger for the fruitful discussions related to the conception and implementation of the presented work as well as for peer reviewing this paper before publication.  Furthermore, we highly appreciate the support of Dominik Rieder, Maximilian Will and Renbo Pan for implementing parts of the developed concepts within their theses.

## 7  References

[1]  *Taxonomy and Definitions for Terms Related to Driving Automation Systems for On-Road Motor Vehicles*, SAE International Std. J3016, 2021,
doi: 10.4271/J3016_202104.

[2]  Hallerbach, S., Xia, Y., Eberle, U., and Koester, F., "Simulation-Based Identification of Critical Scenarios for Cooperative and Automated Vehicles," *SAE Intl. J CAV* 1(2):93-106, 2018, https://doi.org/10.4271/2018-01-1066.

[3]  Jesenski, J. E. Stellet, W. Branz and J. M. Zöllner, "Simulation-Based Methods for Validation of Automated Driving: A Model-Based Analysis and an Overview about Methods for Implementation," 2019 IEEE Intelligent Transportation Systems Conference (ITSC), 2019, pp. 1914-1921,
doi: 10.1109/ITSC.2019.8917072.

[4]  Aparow, A. Choudary, G. Kulandaivelu, T. Webster, J. Dauwels and N. d. Boer, "A Comprehensive Simulation Platform for Testing Autonomous Vehicles in 3D Virtual Environment," 2019 IEEE 5th International Conference on Mechatronics System and Robots (ICMSR), 2019, pp. 115-119,
doi: 10.1109/ICMSR.2019.8835477.

[5]  P. Junietz, W. Wachenfeld, K. Klonecki and H. Winner, "Evaluation of Different Approaches to Address Safety Validation of Automated Driving," 2018 21st International Conference on Intelligent Transportation Systems (ITSC), 2018, pp. 491-496, doi: 10.1109/ITSC.2018.8569959.

[6] T. Duy Son, A. Bhave and H. Van der Auweraer, "Simulation-Based Testing Framework for Autonomous Driving Development," 2019 IEEE International Conference on Mechatronics (ICM), 2019, pp. 576-583, doi: 10.1109/ICMECH.2019.8722847.

[7] Q. Chao, X. Jin, H.-W. Huang, S. Foong, L.-F. Yu and S.-K. Yeung, "Force-based Heterogeneous Traffic Simulation for Autonomous Vehicle Testing," 2019 International Conference on Robotics and Automation (ICRA), 2019, pp. 8298-8304, doi: 10.1109/ICRA.2019.8794430.

[8] H. Bi, T. Mao, Z. Wang and Z. Deng, "A Deep Learning-Based Framework for Intersectional Traffic Simulation and Editing," in IEEE Transactions on Visualization and Computer Graphics, vol. 26, no. 7, pp. 2335-2348, 1 July 2020, doi: 10.1109/TVCG.2018.2889834.

[9] C. Sippl, B. Schwab, P. Kielar and A. Djanatliev, "Distributed Real-Time Traffic Simulation for Autonomous Vehicle Testing in Urban Environments," 2018 21st International Conference on Intelligent Transportation Systems (ITSC), Maui, HI, 2018, pp. 2562-2567, doi: 10.1109/ITSC.2018.8569544.

[10] Statistisches Bundesamt – Verkehrsunfälle. Available at https://www.destatis.de/DE/Themen/Gesellschaft-Umwelt/Verkehrsunfaelle/_inhalt.html (accessed: 03.12.2020)

[11] P. M. Boesch and F. Ciari, "Agent-based simulation of autonomous cars," 2015 American Control Conference (ACC), Chicago, IL, 2015, pp. 2588-2592, doi: 10.1109/ACC.2015.7171123.

[12] Wachenfeld, Walther, and Hermann Winner. "Die Freigabe des autonomen Fahrens." Autonomes Fahren. Springer Vieweg, Berlin, Heidelberg, 2015. 439-464.

[13] *Road Vehicles – Functional Safety*, ISO Std. 26262, 2018.

[14] *Road vehicles — Safety of the intended functionality*, ISO/PAS 21448, 2019.

[15] Wachenfeld, Walther, and Hermann Winner. "The release of autonomous vehicles." *Autonomous driving*. Springer, Berlin, Heidelberg, 2016. 425-449.

[16] PEGASUS Consortium, "PEGASUS method: An overview," 2019, Accessed: June 21, 2021. [Online]. Available: https://www.pegasusprojekt.de/files/tmpl/Pegasus-Abschlussveranstaltung/PEGASUS-Gesamtmethode.pdf

[17] T. Menzel, G. Bagschik and M. Maurer, "Scenarios for Development, Test and Validation of Automated Vehicles," 2018 IEEE Intelligent Vehicles Symposium (IV), 2018, pp. 1821-1827, doi: 10.1109/IVS.2018.8500406.

[18]　E. de Gelder and J. Paardekooper, "Assessment of Automated Driving Systems using real-life scenarios," 2017 IEEE Intelligent Vehicles Symposium (IV), 2017, pp. 589-594, doi: 10.1109/IVS.2017.7995782.

[19]　A. Erdogan et al., "Real- World Maneuver Extraction for Autonomous Vehicle Validation: A Comparative Study," 2019 IEEE Intelligent Vehicles Symposium (IV), 2019, pp. 267-272, doi: 10.1109/IVS.2019.8814254.

[20]　F. Kruber, J. Wurst, E. S. Morales, S. Chakraborty and M. Botsch, "Unsupervised and Supervised Learning with the Random Forest Algorithm for Traffic Scenario Clustering and Classification," 2019 IEEE Intelligent Vehicles Symposium (IV), 2019, pp. 2463-2470, doi: 10.1109/IVS.2019.8813994.

[21]　M. Wood et al., "Safety first for automated driving," 2019. Accessed: June. 20, 2021. [Online]. Available: https://www.daimler.com/dokumente/innovation/sonstiges/safety-first-for-automated-driving.pdf

[22]　C. Neurohr, L. Westhofen, M. Butz, M. H. Bollmann, U. Eberle and R. Galbas, "Criticality Analysis for the Verification and Validation of Automated Vehicles," in IEEE Access, vol. 9, pp. 18016-18041, 2021,
doi: 10.1109/ACCESS.2021.3053159.

[23]　H. Watanabe, T. Maly, J. Wallner, T. Dirndorfer, M. Mai, and G. Prokop,"Methodology of scenario clustering for predictive safety functions," in 9. Tagung Automatisiertes Fahren. Munich, Germany: Technical Univ. of Munich, 2019.

[24]　S. Ulbrich, T. Menzel, A. Reschka, F. Schuldt and M. Maurer, "Defining and Substantiating the Terms Scene, Situation, and Scenario for Automated Driving," 2015 IEEE 18th International Conference on Intelligent Transportation Systems, 2015, pp. 982-988, doi: 10.1109/ITSC.2015.164.

[25]　Sippl, C. S. Identifikation relevanter Verkehrssituationen für die szenarienbasierte Entwicklung automatisierter Fahrfunktionen (Doctoral dissertation, Friedrich-Alexander-Universität Erlangen-Nürnberg (FAU)), 2020.

[26]　J. Bock, R. Krajewski, T. Moers, S. Runde, L. Vater and L. Eckstein, "The inD Dataset: A Drone Dataset of Naturalistic Road User Trajectories at German Intersections," 2020 IEEE Intelligent Vehicles Symposium (IV), 2020, pp. 1929-1934, doi: 10.1109/IV47402.2020.9304839.

[27]　Sander, U., & Lubbe, N. (2018). The potential of clustering methods to define intersection test scenarios: Assessing real-life performance of AEB. *Accident Analysis & Prevention*, *113*, 1-11.

[28]　M. Barbier, C. Laugier, O. Simonin and J. Ibañez-Guzmán, "Classification of drivers manoeuvre for road intersection crossing with synthethic and real-data," 2017 IEEE Intelligent Vehicles Symposium (IV), 2017, pp. 224-230, doi: 10.1109/IVS.2017.7995724.

[29] Hartjen, L., Philipp, R., Schuldt, F., Friedrich, B., & Howar, F. (2019). Classification of driving maneuvers in urban traffic for parametrization of test scenarios. In *9. Tagung Automatisiertes Fahren*.

[30] King, C., Braun, T., Braess, C., Langner, J., & Sax, E. (2021). Capturing the Variety of Urban Logical Scenarios from Bird-view Trajectories. In *VEHITS* (pp. 471-480).

[31] N. Weber, D. Frerichs and U. Eberle, "A simulation-based, statistical approach for the derivation of concrete scenarios for the release of highly automated driving functions," AmE 2020 - Automotive meets Electronics; 11th GMM-Symposium, 2020, pp. 1-6.

[32] T. Duy Son, A. Bhave and H. Van der Auweraer, "Simulation-Based Testing Framework for Autonomous Driving Development," 2019 IEEE International Conference on Mechatronics (ICM), 2019, pp. 576-583, doi: 10.1109/ICMECH.2019.8722847.

[33] Calvert, S. C., & van Arem, B. (2020). A generic multi-level framework for microscopic traffic simulation with automated vehicles in mixed traffic. *Transportation Research Part C: Emerging Technologies*, *110*, 291-311.

[34] Cantas, M. and Guvenc, L., "Customized Co-Simulation Environment for Autonomous Driving Algorithm Development and Evaluation," SAE Technical Paper 2021-01-0111, 2021, https://doi.org/10.4271/2021-01-0111.

[35] D. Nalic, A. Eichberger, G. Hanzl, M. Fellendorf and B. Rogic, "Development of a Co-Simulation Framework for Systematic Generation of Scenarios for Testing and Validation of Automated Driving Systems*," 2019 IEEE Intelligent Transportation Systems Conference (ITSC), 2019, pp. 1895-1901, doi: 10.1109/ITSC.2019.8916839.

[36] German funded project of PEGASUS family "SET Level – Simulation-based development and testing of automated driving," Accessed: June 20, 2021. [Online]. Available: https://setlevel.de/en

[37] Wachenfeld, W., & Winner, H. (2015). Virtual assessment of automation in field operation a new runtime validation method. In *10. Workshop Fahrerassistenzsysteme* (p. 161).

[38] C. Wang and H. Winner, "Overcoming Challenges of Validation Automated Driving and Identification of Critical Scenarios," 2019 IEEE Intelligent Transportation Systems Conference (ITSC), 2019, pp. 2639-2644, doi: 10.1109/ITSC.2019.8917045.

[39] Koenig, A., Witzlsperger, K., Leutwiler, F., & Hohmann, S. (2018). Overview of HAD validation and passive HAD as a concept for validating highly automated cars. *at-Automatisierungstechnik*, *66*(2), 132-145.

[40] Nalic D, Pandurevic A, Eichberger A, Fellendorf M, Rogic B. Software Framework for Testing of Automated Driving Systems in the Traffic Environment of Vissim. *Energies.* 2021; 14(11):3135. https://doi.org/10.3390/en14113135

[41] German funded project of PEGASUS family "VVM - Verification and validation methods for automated vehicles in urban environments," Accessed: June 21, 2021. [Online]. Available: https://www.vvm-projekt.de/en/

[42] SET Level research project, "Simulation Use Case 1 – A closed-loop traffic simulation for criticality analysis," Accessed: June 23, 2021. [Online]. Available: https://www.youtube.com/watch?v=iJNWYZNtpwM

[43] F. Schuldt, "Ein Beitrag für den methodischen Test von automatisierten Fahrfunktionen mit Hilfe von virtuellen Umgebungen," Ph.D. dissertation, Technische Universität Braunschweig, Braunschweig, Germany, 2017.

[44] G. Bagschik, T. Menzel and M. Maurer, "Ontology based Scene Creation for the Development of Automated Vehicles," 2018 IEEE Intelligent Vehicles Symposium (IV), 2018, pp. 1813-1820, doi: 10.1109/IVS.2018.8500632.

[45] Bock, J., Krajewski, R., Eckstein, L., Klimke, J., Sauerbier, J., & Zlocki, A. (2018). Data basis for scenario-based validation of HAD on highways. In *27th Aachen colloquium automobile and engine technology* (pp. 8-10).

[46] Weber, H., Bock, J., Klimke, J., Roesener, C., Hiller, J., Krajewski, R., ... & Eckstein, L. (2019). A framework for definition of logical scenarios for safety assurance of automated driving. Traffic injury prevention, 20(sup1), S65-S70.

[47] M. Scholtes et al., "6-Layer Model for a Structured Description and Categorization of Urban Traffic and Environment," in IEEE Access, vol. 9, pp. 59131-59147, 2021, doi: 10.1109/ACCESS.2021.3072739.

[48] S. Riedmaier, T. Ponn, D. Ludwig, B. Schick and F. Diermeyer, "Survey on Scenario-Based Safety Assessment of Automated Vehicles," in IEEE Access, vol. 8, pp. 87456-87477, 2020, doi: 10.1109/ACCESS.2020.2993730.

[49] Nalic, D., Mihalj, T., Bäumler, M., Lehmann, M., Eichberger, A., & Bernsteiner, S. (2020). *Scenario Based Testing of Automated Driving Systems: A Literature Survey*. 1. Paper presented at FISITA Web Congress 2020, Virtuell, Czech Republic.

[50] Winner, H.: Einrichtung zum Bereitstellen von Signalen in einem Kraftfahrzeug. Patent Application DE000010102771A1 (2001)

[51] Reschka, A., Rieken, J., Maurer, M.: Entwicklungsprozess von Kollisionsschutzsystemen für Frontkollisionen: Systeme zur Warnung, zur Unfallschwereminderung und zur Verhinderung. In: Winner, H., Hakuli, S., Lotz, F., Singer, C. (Hrsg.) Handbuch Fahrerassistenzsysteme, 3rd edn., pp. 913–935. Vieweg-Teubner-Verlag (2015)

Preprint – 30th Aachen Colloquium Sustainable Mobility 2021    35[52]  Junietz, P., Wachenfeld, W., Schönemann, V., Domhardt, K., Tribelhorn, W., & Winner, H. (2019). Gaining Knowledge on Automated Driving's Safety—The Risk-Free VAAFO Tool. In *Control Strategies for Advanced Driver Assistance Systems and Autonomous Driving Functions* (pp. 47-65). Springer, Cham.

[53]  C. Wang, F. Xiong and H. Winner, "Reduction of Uncertainties for Safety Assessment of Automated Driving Under Parallel Simulations," in IEEE Transactions on Intelligent Vehicles, vol. 6, no. 1, pp. 110-120, March 2021, doi: 10.1109/TIV.2020.2987437.

[54]  Eberle, U., Hallerbach, S.: Verfahren zum Trainieren wenigstens eines Algorithmus für ein Steuergerät eines Kraftfahrzeugs, Computerprogrammprodukt, Kraftfahrzeug sowie System. Patent Application DE 10 2019 206 908 A1 (2020)

[55]  Steimle, M., Menzel, T., & Maurer, M. (2021). Towards a Consistent Terminology for Scenario-Based Development and Test Approaches for Automated Vehicles: A Proposal for a Structuring Framework, a Basic Vocabulary, and its Application. *arXiv preprint arXiv:2104.09097*.

[56]  B. Strasser, "Vernetzung von Test- und Simulationsmethoden für die Entwicklung von Fahrerassistenzsystemen," Ph.D. dissertation, Technische Universität München, Munich, Germany, 2012.

[57]  Hallerbach, S. (2020). *Simulation-based testing of cooperative and automated vehicles* (Doctoral dissertation, Universität Oldenburg).

[58]  Kober, C. (2019). *Stochastische Verkehrsflusssimulation auf Basis von Fahrerverhaltensmodellen zur Absicherung automatisierter Fahrfunktionen*. Springer-Verlag. ISBN: 978-3-658-25251-9

[59]  Dallmeyer, J. (2014). Simulation des Straßenverkehrs in der Großstadt: das Mit- und Gegeneinander verschiedener Verkehrsteilnehmertypen. Springer-Verlag. ISBN: 978-3-658-05207-2

[60]  Becker, D., Klimke, J., & Eckstein, L. (2021). Agent Model for the Closed-loop Simulation of Traffic Scenarios. *ATZelectronics worldwide*, *16*(5), 40-43.

[61]  Association for Standardization of Automation and Measuring Systems (ASAM), "ASAM SIM:GUIDE - Standardization for Highly Automated Driving," 2021, Accessed: June 23, 2021. [Online]. Available: https://www.asam.net/index.php?eID=dumpFile&t=f&f=4297&token=cbeec18c4770d0ff88e3d3c539c13a32abed6300

[62]  Association for Standardization of Automation and Measuring Systems (ASAM), Accessed: June 23, 2021. [Online]. Available: https://www.asam.net

[63]  Functional Mock-up Interface (FMI), Accessed: June 23, 2021. [Online]. Available: https://fmi-standard.org/

36    Preprint – 30th Aachen Colloquium Sustainable Mobility 2021
[64] F. Poggenhans et al., "Lanelet2: A high-definition map framework for the future of automated driving," 2018 21st International Conference on Intelligent Transportation Systems (ITSC), 2018, pp. 1672-1679, doi: 10.1109/ITSC.2018.8569929.

[65] P. Cooper, "Experience with traffic conflicts in canada with emphasis on "post encroachment time" techniques", in International calibration study of traffic conflict techniques, Springer, 1984, pp. 75–96.

[66] Songchitruksa, P., & Tarko, A. P. (2006). Practical method for estimating frequency of right-angle collisions at traffic signals. *Transportation research record*, *1953*(1), 89-97.

[67] Archer, J. (2005). *Indicators for traffic safety assessment and prediction and their application in micro-simulation modelling: A study of urban and suburban intersections* (Doctoral dissertation, KTH).

[68] Hydén, C. (1996). Traffic conflicts technique: state-of-the-art. *Traffic safety work with video processing*, *37*, 3-14.

[69] Mahmud, S. S., Ferreira, L., Hoque, M. S., & Tavassoli, A. (2017). Application of proximal surrogate indicators for safety evaluation: A review of recent developments and research needs. *IATSS research*, *41*(4), 153-163.

[70] Wei, L., Li, Z., Gong, J., Gong, C., & Li, J. (2021). Autonomous Driving Strategies at Intersections: Scenarios, State-of-the-Art, and Future Outlooks. *arXiv preprint arXiv:2106.13052*.

[71] M. M. Mubasher and J. Syed Waqar ul Qounain, "Systematic literature review of vehicular traffic flow simulators," 2015 International Conference on Open Source Software Computing (OSSCOM), 2015, pp. 1-6, doi: 10.1109/OSSCOM.2015.7372687.

[72] M. Treiber und A. Kesting, *Verkehrsdynamik und -simulation: Daten, Modelle und Anwendungen der Verkehrsflussdynamik*, 1st ed. 2010, Ser. Springer-Lehrbuch. Berlin, Heidelberg und Cham: Springer Berlin Heidelberg und Springer International Publishing AG, 2010, isbn: 9783642052286. doi: 10.1007/978-3-642-05228-6.

[73] PTV Vissim, Accessed: July 1, 2021. [Online]. Available: https://www.ptvgroup.com/de/loesungen/produkte/ptv-vissim/

[74] P. A. Lopez et al., "Microscopic Traffic Simulation using SUMO," 2018 21st International Conference on Intelligent Transportation Systems (ITSC), 2018, pp. 2575-2582, doi: 10.1109/ITSC.2018.8569938.

[75] Aimsun, Accessed: July 1, 2021. [Online]. Available: https://www.aimsun.com/

[76] CoEXist Project, Accessed: July 1, 2021. [Online]. Available: https://www.h2020-coexist.eu/

[77] VeriCAV Project, Accessed: July 1, 2021. [Online]. Available: https://vericav-project.co.uk/

[78] Eclipse SUMO - Documentation, Accessed: July 4, 2021. [Online]. Available: https://sumo.dlr.de/docs/index.html

[79] PTV AG, Ed., PTV Vissim 2020 User Manual, 2020.00-07, Haid-und-Neu-Str. 15, 76131Karlsruhe, Germany, Apr. 2020

[80] Helbing, D., & Molnar, P. (1995). Social force model for pedestrian dynamics. *Physical review E*, *51*(5), 4282.

[81] Huang, W., Fellendorf, M., & Schönauer, R. (2011). Social force based vehicle model for 2-dimensional spaces. In *91st Annual Meeting of the Transportation Research Board. Washington, DC, USA*.